\documentclass{article} 
\usepackage{iclr2021_conference,times}

\usepackage{amsmath,amsfonts,bm}

\def\eqref#1{equation~\ref{#1}}

\def\1{\bm{1}}

\DeclareMathAlphabet{\mathsfit}{\encodingdefault}{\sfdefault}{m}{sl}
\SetMathAlphabet{\mathsfit}{bold}{\encodingdefault}{\sfdefault}{bx}{n}

\usepackage{hyperref}
\usepackage{graphicx}
\usepackage{url}
\usepackage{flushend}

\usepackage{url}

\title{Vaccination Worldwide: Strategies, Distribution and Challenges}

\author{\\\textbf{Chirag Samal}\textsuperscript{1}, \textbf{Kasia Jakimowicz}\textsuperscript{1,2}, \textbf{Krishnendu Dasgupta}\textsuperscript{1},  \textbf {Aniket Vashishtha}\textsuperscript{1},  \textbf {Francisco O.}\textsuperscript{1, 3}, \\ \textbf {Arunakiry Natarajan}\textsuperscript{1},  \textbf{Haris Nazir}\textsuperscript{1}, \textbf{Alluri Siddhartha Varma}\textsuperscript{1}, \textbf {Tejal Dahake}\textsuperscript{1}, \textbf {Amitesh Anand} \\ \textbf{Pandey}\textsuperscript{1}, \textbf {Ishaan Singh}\textsuperscript{1}, \textbf {John Sangyeob Kim}\textsuperscript{1}, \textbf{Mehrab} \textbf{Singh Gill}\textsuperscript{1}, \textbf{Saurish Srivastava}\textsuperscript{1}, \textbf{Orna}\\ \textbf{ Mukhopadhyay}\textsuperscript{1}, \textbf{Parth Patwa}\textsuperscript{1},\textbf{Qamil Mirza}\textsuperscript{1}, \textbf{Sualeha Irshad}\textsuperscript{1},
\textbf{Sheshank Shankar}\textsuperscript{1,2}, \textbf{Rohan} \\\textbf{Iyer}\textsuperscript{1}, \textbf{Rohan Sukumaran}\textsuperscript{1}, \textbf{Ashley} \textbf{Mehra}\textsuperscript{1}, \textbf{Anshuman Sharma}\textsuperscript{1}, \textbf{Abhishek Singh}\textsuperscript{2}, \textbf{Maurizio}\\ \textbf{Arseni}\textsuperscript{1}, \textbf{Sethuraman T V}\textsuperscript{1}, \textbf{Saras Agrawal}\textsuperscript{1},
\textbf{Vivek Sharma}\textsuperscript{1,2,3},
\textbf{Ramesh Raskar}\textsuperscript{1,2} \\\\

\textsuperscript{1} PathCheck Foundation, 02139 Cambridge, USA.\\
\textsuperscript{2} MIT Media Lab, 02139 Cambridge, USA.\\
\textsuperscript{3} Harvard Medical School, USA.\\
\texttt{raskar@media.mit.edu} }

\iclrfinalcopy 
\begin{document}

\maketitle

\begin{abstract}
  The Coronavirus 2019 (Covid-19) pandemic caused by the SARS-CoV-2 virus represents an unprecedented crisis for our planet. It is a bane of the über connected world that we live in that this virus has affected almost all countries and caused mortality and economic upheaval at a scale whose effects are going to be felt for generations to come. While we can all be buoyed at the pace at which vaccines have been developed and brought to market, there are still challenges ahead for all countries to get their populations vaccinated equitably and effectively. This paper provides an overview of ongoing immunization efforts in various countries. In this early draft, we have identified a few key factors that we use to review different countries' current COVID-19 immunization strategies and their strengths and draw conclusions so that policymakers worldwide can learn from them. Our paper focuses on processes related to vaccine approval, allocation and prioritization, distribution strategies, population to vaccine ratio, vaccination governance, accessibility and use of digital solutions, and government policies. The statistics and numbers are dated as per the draft date [June 24th,  2021].
\end{abstract}

\section{Introduction}
 Currently, at least seven different vaccines have been rolled out across all countries. Scientific knowledge to make well-informed decisions for vaccine distribution strategy is highly important \citep{jeyanathan2020immunological}. Countries worldwide currently approve different vaccines and prepare their vaccine roll-out plans, working on the ongoing vaccination effort based on their priority groups. Various vaccination efforts have their pros and cons \citep{mills2021challenges},\citep{baechallenges},\citep{sallam2021covid},  and we summarize the study of vaccine distribution and administration in different countries \cite{kim2021looking}. \\
 In this early draft, the paper focuses on five countries and contributing factors such as approved vaccines, priority groups, vaccination administration, and distribution.\\

The paper examines the following factors of vaccine supply affecting the last-mile administration across the world: 
\begin{enumerate}
    \item Prioritization: This focuses on the prioritization schemes for the vaccine administered by the government and health authorities. We include a timeline where they have been made available as per the country's public records. Certain population groups have been identified as to be vaccinated first, as they are at greater risk of exposure or serious consequences of COVID-19 \citep{persad2020fairly}, \citep{national2020framework}, \citep{wang2020global}.
    \item Approved vaccines: This section includes information on vaccines approved for distribution in the country.
    \item Vaccine Administration and Distribution : This section contains necessary strategies taken by the country for equitable distribution, logistics and regulations, and timelines \citep{ismail2020framework}.
    \item Population to Vaccine ratio : This ratio is calculated as per dosage per 100 people on any given date as recorded by the national vaccination control board. Depending on the individual dosing schedule, this is counted as a single dose and does not reflect the total number of people vaccinated.
    \item Vaccination Governance: This section mentions the governing bodies of the country for the vaccination process.
    \item Accessibility and Digital Solutions : This section covers accessibility of all vaccine-related information services, scheduling service, the authenticity of vaccine coupons\citep{bae2021safepaths} and overall vaccination insights available, and intermediary data collection through any digital \citep{bae2021mobile} or offline-based medium.
    \item Government Strategies and Decisions : This largely focuses on government initiatives and decisions that were imposed as a directive or policy for vaccination \citep{piraveenan2020optimal}, \citep{acuna2020covid}, \citep{valiati2021mitigation}.
\end{enumerate}
\begin{figure*}
    \centering
    \includegraphics[width = 14cm]{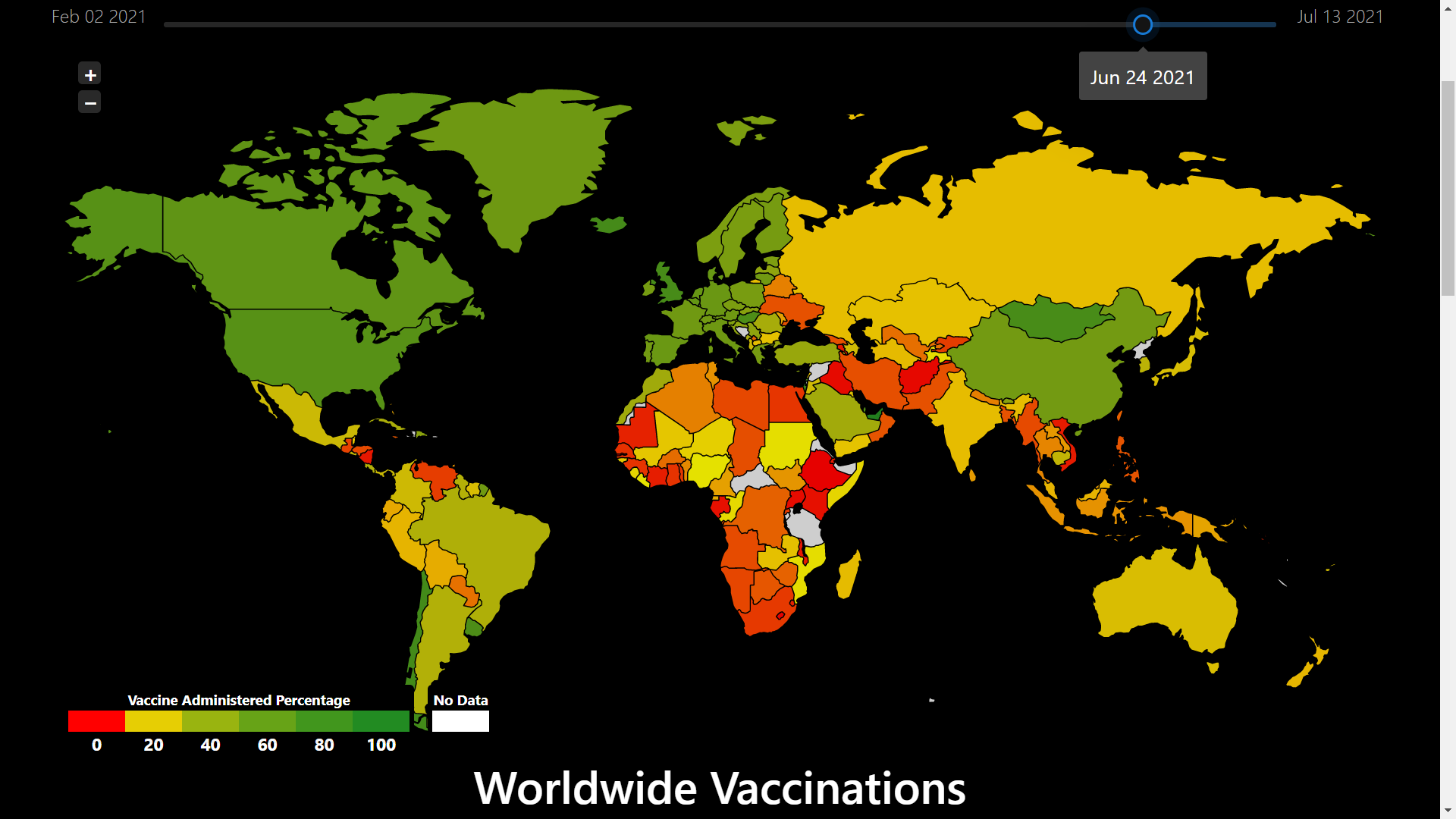}
    \vspace{-10pt}
    \caption{Vaccinations administered by Country. The map depicts the total number of COVID-19 vaccine doses administered in each region. Note that this amount represents a single dose depending on the particular dosing schedule and does not represent the total number of people vaccinated (e.g., people will receive multiple doses). Official data collected from the Pathcheck Foundation's Vaccination Dashboard (https://vax.uab.edu/) - Last updated June 24, 2021.}
    \label{fig:1}
\end{figure*}
\begin{figure*}
    \centering
    \vspace{-20pt}
    \includegraphics[width = 14cm, ,keepaspectratio]{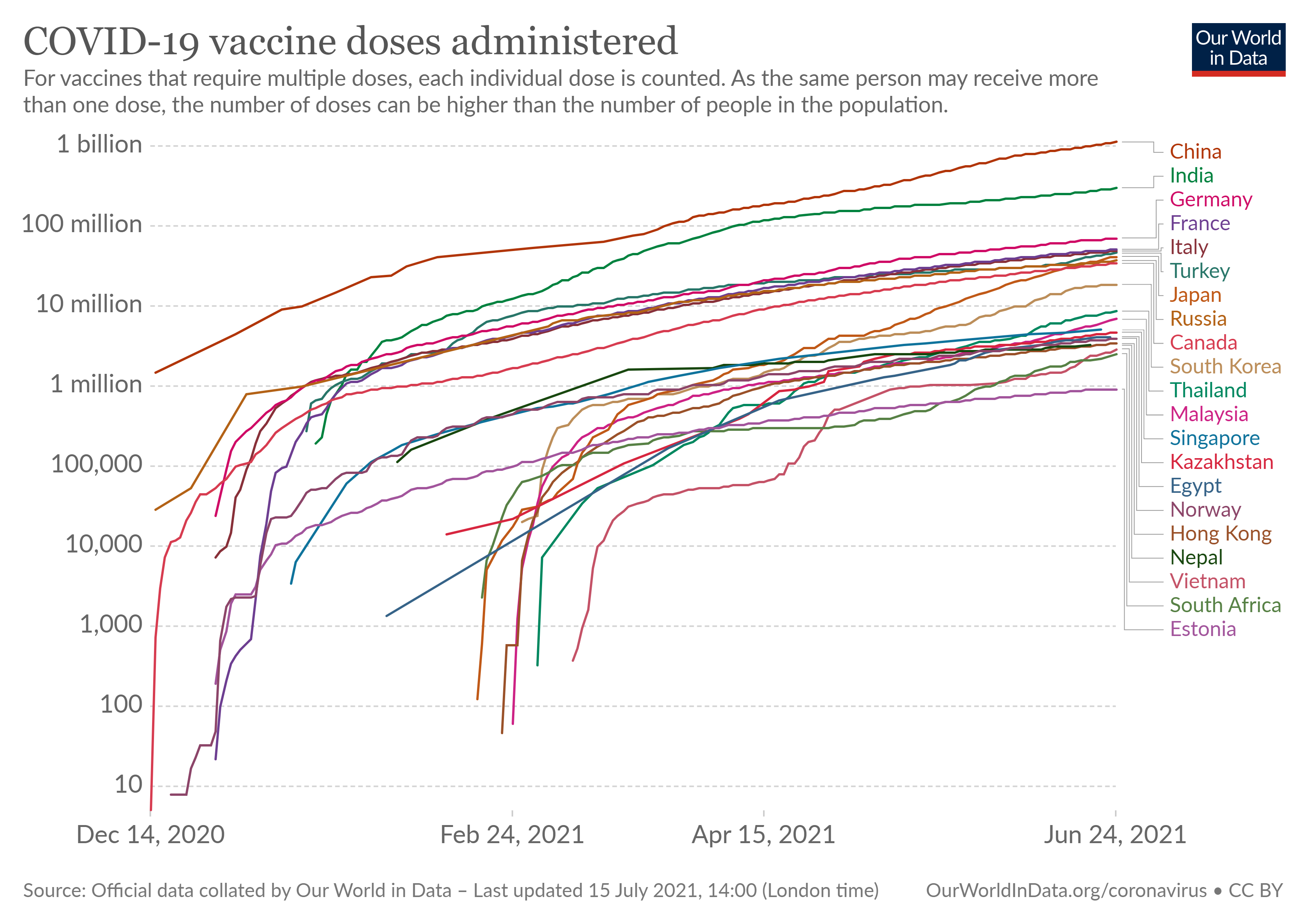}
    \vspace{-20pt}
    \caption{Total COVID-19 doses administered by countries. The graph depicts the total number of COVID-19 vaccine doses given out by different countries studied in this work. Note that this amount represents a single dose depending on the particular dosing schedule and does not represent the total number of people vaccinated (e.g., people are given multiple doses). Official data by Our World in Data (https://ourworldindata.org/covid-vaccinations) - Last updated June 24, 2021.}
    \label{fig:2}
\end{figure*}

\section{Countries}
\subsection{India}

\subsubsection{Prioritization}
The schedule for the roll-out of vaccination in these three priority groups is adjusted according to the availability of the vaccine. It is planned to vaccinate nearly 30,000 people in the first phase of the vaccination campaign~\citep{india5}:
\begin{enumerate}
    \item Health workers (HCW): health care providers and workers in health care facilities.
    \item Frontline Workers (FLWs): state and central police personnel, armed forces, custody and accommodation, prison staff, disaster management, and civil defense volunteers, municipal volunteers workers, and finance officials engaged in the containment, surveillance, and related activities of COVID-19.
    \item population $\geq$ 60 years and $<$ 45 years with co-morbidities such as diabetes, hypertension, cancer, lung disease, etc.
    \item Population above age 45 years
\end{enumerate}

\subsubsection{Approved Vaccines}
Two vaccines have been approved so far:
\begin{itemize}
    \item Covaxin (Bharat Biotech) \citep{ella2020safety}
    \item Covishield (Oxford-AstraZeneca) \citep{knoll2021oxford}
\end{itemize}

\subsubsection{Vaccine Administration and Distribution}
Vaccination is carried out in fixed and nearby locations. Fixed locations with public health facilities such as primary health centers and private health facilities with more than 100 medical professionals will be used for vaccination sites. Besides, outreach facilities such as schools, colleges, community halls, Panchayat Bhawans, wedding venues, etc., can also be used. Mobile teams can be used in some inaccessible areas or areas with a migrant population. Immunization days and times are planned and determined in advance. Vaccination and all auxiliary workers and beneficiaries are assigned to a specific location. Only one type of vaccine will be given on a given day, at a given vaccination site, and each beneficiary will receive a second dose of the same vaccine.

\textbf{Vaccination Rate}
\begin{itemize}
    \item 3.8\% fully vaccinated till now
    \item 300 million vaccine doses have been administered
    \item 52.4 million people fully vaccinated
    \item Population to Vaccine ratio: 21 COVID-19 vaccine doses administered per 100 people as recorded on June 24, 2021. 3.8\% of the total population of the country have received all doses prescribed by the vaccination protocol \cite{Ourworldindata}
\end{itemize}

\textbf{Vaccination Governance}: India has set up an apex body called NEGVAC (National Expert Group On Vaccine Administration For COVID-19) to provide guidance on all issues related to vaccine trial and selection, distribution of vaccine, developing delivery mechanisms, priority allocation to population groups, vaccine safety surveillance, mass communication, etc. Key Ministries and Institutions with on-the-ground know-how are cooperating and collaborating at the national level.

\subsubsection{Accessibility and Digital Solutions}
Co-WIN (COVID-19 Vaccine Intelligence Network) was developed in India as a robust cloud-based IT solution for the preparation, implementation, monitoring, and evaluation of the COVID-19 immunization program \citep {sahoo2021concepts}. It is a holistic approach that uses the entire public health system, from the national level down to the vaccination doctor level. The system allows the registration of beneficiaries, planning units, vaccination sites, planning meetings, and the implementation of the vaccination process. The real-time Co-WIN system is used to track beneficiaries and vaccinators at the national, state, and district levels. Administrators at the national, state, and district levels can use the website www.cowin.gov.in to manage and distribute immunization-related materials, identify immunization sites, and register either by bulk download or by self-registration of beneficiaries. The www.app.cowin.gov.in application allows registering individual beneficiaries and registering successful vaccinations at vaccination sites~\citep{india4}.

Vaccines are sent to the vaccination sites in suitable containers on the same day in order to meet the temperature requirements throughout the cold chain, which will be monitored by the Ministry of Health. Text messages or self-created emails are sent to all beneficiaries and the vaccination and support staff regarding the date, time, and place of the vaccination appointment. The vaccination module of the Co-WIN system is used by the vaccination team to verify and authenticate the beneficiaries using Aadhar contact details or another approved photo ID. Once authenticated, the vaccine is administered and the status is updated in the system. The recipient will then receive an SMS with the details of a link indicating the date and time of the next vaccination. The Co-WIN system tracks every vaccine dose (name, manufacturer, batch number, date of manufacture) to every vaccination site and recipient~\citep{india5}.

\subsubsection{Government Strategies and Decisions}
Various preparatory activities were carried out in the government advisory body for the COVID-19 vaccine roll-out in collaboration with States/ Union territories. A multi-level coordination system was set up for vaccinating priority groups, which includes strengthening the cold chain infrastructure, additional vaccinators across the country, a communication strategy to address the safety and effectiveness of the vaccine among the masses. Co-WIN, a digital platform for COVID-19 Vaccination, will be used to track the beneficiaries and vaccines on a real-time basis.~\citep{india5}.

\subsubsection{Money Spent}
According to data provided to the Rajya Sabha by the Ministry of Health and Family Welfare, the Centre spent \$191 million on COVID-19 vaccines from the Serum Institute of India and Indian producer Bharat Biotech between early January and mid-March (MoHFW).This is a preliminary indication of the direct cost of the COVID-19 immunization program to the exchequer.

Additionally, the government has earmarked \$65 million to meet the operation and maintenance costs incurred by the state governments in this fiscal. So far, the Centre has released \$17 million to states. \cite{india6}

\subsubsection{Variant Info}
In numerous states with high case numbers, samples bearing the "double mutant," or B.1.617 variety, have been discovered.

Maharashtra, Karnataka, West Bengal, Gujarat, and Chhattisgarh were among the states where the B.1.617 variation was found. \cite{india7}

\subsection{Japan}

\subsubsection{Prioritization}
Healthcare workers, the elderly, and those with pre-existing health conditions.

DISTRIBUTION TIMELINE~\citep{Japan2}:
      
\begin{itemize}
    \item In February, 10,000 frontline healthcare employees will be vaccinated; 
    \item Mid-March: 3 million general healthcare workers will be vaccinated.
    \item Late March to early April: 30-40 million elderly people will be vaccinated.
    \item Citizens with pre-existing ailments will be eligible to get vaccinated starting in April.
    \item End-June: The government hopes to provide enough doses for the whole population, but the exact date has yet to be determined.
\end{itemize}

\subsubsection{Approved Vaccines}
Three vaccines have been approved so far:
\begin{itemize}
    \item Pfizer-BioNTech \citep{polack2020safety}
    \item Oxford-AstraZeneca \citep{knoll2021oxford}
    \item Moderna \citep{baden2021efficacy}
\end{itemize}

\subsubsection{Vaccine Administration and Distribution}
The government is developing a system to streamline and share information on vaccine stocks at all medical institutions. The responsibility of promoting the vaccine program will fall onto the local governments. Once the national immunization campaign launches, the public will have to enroll in a reservation system in order to receive a vaccine. Initially, they will register on a vaccination website or through a call center, and they will then obtain a vaccine voucher that will enable them to receive a shot. The vaccine coupon will also be proof of vaccination for each immunized citizen. Currently, Japan has not released much more information on the details of its vaccine distribution plan.

\textbf{Vaccination Rate}
\begin{itemize}
    \item 35.7 million doses administered
    \item 11 million people fully vaccinated
    \item 8.7\% population fully vaccinated
    \item Population to Vaccine ratio: 28.3 COVID-19 vaccine doses administered per 100 people as recorded on June 24, 2021 \cite{Ourworldindata}
    \item In the midst of a fourth wave of illnesses, Japan has hastened the distribution of vaccinations with the goal of vaccinating all of the elderly by the end of July \cite{Japan20}
\end{itemize}

\textbf{Vaccination Governance}: Vaccine administration in Japan is governed by the Japanese Health Ministry.

\subsubsection{Accessibility and Digital Solutions}
A reservation system will be developed for the general public to enroll for vaccination. Japan plans to use medical facilities, public health centers, gyms, event venues, and shopping malls as vaccination sites and is also working to re-purpose non-medical facilities as vaccination sites~\citep{Japan5}.

\subsubsection{Government Strategies and Decisions}
Japan has already passed legislation to provide free COVID-19 vaccines to all inhabitants, and has set aside approximately \$ 6.32 billion from the emergency budget for COVID-19 vaccines~\citep{Japan16}.
Japan intends to collect data from persons infected with the new coronavirus, even if they have been vaccinated, to see how effective the vaccines are at preventing the virus from spreading.
The new system will allow authorities to view vaccination records for coronavirus patients, e.g., what vaccines they received from the companies and whether they received a single or double dose, depending on the sources.

In Japan, vaccinations are likely to begin in late February. According to the sources, the Ministry of Health would build a method to collect vaccination records from all afflicted persons by including checkboxes on a document that doctors must send to public health centres when coronavirus infection is confirmed.~\citep{Japan14}.

\subsubsection{Money spent}
According to the Ministry of Finance, the Japanese government approved a budget of \$6.32 billion from its emergency budget reserve to secure COVID-19 vaccinations. The Japanese government has stated that it is willing to cover the expense of supplying vaccinations to the whole population in order to carry out a thorough pandemic inoculation campaign. According to reports, the government is planning to create funds to pay for vaccine-related adverse effects that might be common. \cite{Japan17}

\subsubsection{Variant info}
Around 30 cases of variations from the United Kingdom and South Africa were previously documented in Japan. Experts are concerned that the mutations appear to be spreading more quickly.

In Japan, a new variation has arisen, dubbed 'Eek' by scientists. According to NHK, a Japanese public station, the new E484K mutation, which is known to reduce vaccination protection, was found in nearly 70\% of coronavirus illness patients at a Tokyo hospital last month. \cite{Japan18} \cite{Japan19}

\subsection{Canada}

\subsubsection{Prioritization}
 People at high risk of serious illness and death from COVID-19:
 \begin{itemize}
    \item Advanced age.
    \item People with high-risk diseases. 
    \item Healthcare, personal-care and nursing staff
    \item Family members of people at higher risk of serious illness and death from COVID-19
    \item Police, firefighters, Grocery stores employees~\citep{canada2}.
\end{itemize}
Distribution Timeline~\citep{canada3}:
\begin{enumerate}
  \item Phase I: December 2020 to February 2021 | Hospital Health care workers.
  \item Phase II: February to March 2021 | Seniors over the age of 80, hospital personnel, neighborhood general practitioners (GPs), and medical professionals who were not immunized during Phase 1.
  \item Phase III: April to June 2021 | People aged 60 to 79.
  \item Phase IV: July to September 2021 | People aged 18 to 59.
\end{enumerate}

\subsubsection{Approved Vaccines}
Two vaccines have been approved so far ~\citep{canada3}: 
\begin{itemize}
    \item Pfizer-BioNTech \citep{polack2020safety}
    \item Moderna \citep{baden2021efficacy}
\end{itemize}

\subsubsection{Vaccine Administration and Distribution}
Canada has a healthcare system that is universal. The Canadian healthcare system is decentralized, with provinces and territories administering it.
 In Canada, the military is playing a central role in vaccine distribution; the government is sending shipments to all ten provinces. Provinces receive an allocated number of doses of vaccine on a weekly basis.
FedEx Express Canada and Innomar Strategies Inc. have been awarded a contract by the Canadian government to provide an end-to-end logistics solution for COVID-19 vaccines.
The contract will support vaccine distribution across. To facilitate fast and effective vaccine distribution throughout Canada, Canada and provide a range of facilities to support storage and transportation to national, territorial, and Indigenous partners.

\textbf{Vaccination Rate}
\begin{itemize}
    \item 33 million total doses given
    \item 8.2 million people fully vaccinated
    \item 21.9\% fully vaccinated
    \item Population to Vaccine ratio: 88.99 COVID-19 vaccine doses administered per 100 people as dated June 24, 2021 \cite{Ourworldindata}
\end{itemize}

\textbf{Vaccination Governance}: The National Advisory Committee on Immunization (NACI) is an autonomous, ongoing, and timely medical, scientific, and public health advisory body that gives the Public Health Agency of Canada (PHAC) independent, ongoing, and timely medical, scientific, and public health advice on PHAC vaccines. ~\citep{canada1}, \citep{ismail2010canada}.

\subsubsection{Accessibility and Digital Solutions}
Public Services and Procurement Canada (PSPC) uses a service provider to add additional functions to the PHAC. This Enhanced National Vaccine Management Information Platform (NVMIP) is a well-developed and ready-to-use monitoring and recording computer system which will manage the use, administration, and reporting of vaccines as the volume of deliveries increases.

People will receive a paper and digital copy of their immunization record cards. The immunization record will be also be stored in the online provincial database, accessible to the individual, public health, and your doctor~\citep{canada4}.

\subsubsection{Government Strategies and Decisions}
The Government of Canada is providing approximately \$ 220 million so that the facility can purchase 15 million doses of vaccine for all Canadians. The COVID-19 vaccine will be free to anyone living in Canada who is eligible. It will then be available to anyone in Canada who is recommended by federal, provincial, and territorial health authorities to be vaccinated.

\subsubsection{Money spent}
Imports in this category fell by more than half (-55\%) from December to January in 2018 and 2019. Imports grew by 21.4 percent to \$56 million this year, a 73.7 percent rise over the previous year.
In January, Canada's imports of COVID-19 vaccinations were anticipated to be over \$24 million. Imports of vaccines for human medicine other than influenza would have declined in January if the COVID-19 vaccines had not been imported, and would have been on level with imports in January 2020. \cite{canada5}

\begin{itemize}
    \item Both Moderna and Pfizer supplied 695,275 doses to Canada in January. On a per-dose basis, this means the Canadian government spent \$34.41 on each dosage item
    \item According to a recent research based on StatCan data, the federal government spent \$16 million in December to purchase the initial shipments of 423,900 Pfizer and Moderna pharmaceuticals, which cost an average of \$37.74 per dose.
\end{itemize}
\cite{canada6}

\subsubsection{Variant info}
The highly contagious B.1.1.7 strain of SARS-CoV-2, which was initially discovered in the UK in early December, is to blame for the rising number of cases in numerous countries, including India and Canada. \cite{canada7}

The Canadian province of British Columbia curbed an outbreak of the P.1 SARS-CoV-2 variant, which was believed to have originated in Brazil through a combination of targeted vaccinations and physical-distancing rules. Health experts widely regarded these moves as lessons in social distancing and controlling the pandemic. \cite{canada8}

\subsection{China}

\subsubsection{Prioritization}
China's top priority groups are doctors, hotel workers, border control personnel, food storage and transportation workers, and travelers.
Nine critical groups of people between the ages of 18 and 59 were identified as being prioritized:~\citep{china8}:
\begin{itemize}
    \item Transport workers
    \item Social workers
    \item Custom inspection officers who deal with imported cold chain products
    \item The guards at the entrance to the ports
    \item International and national transport officials
    \item People who have to travel abroad for professional or personal reasons
    \item Medical Staff
    \item Employees of government agencies
    \item Employees of public enterprises
    \item Members of the People's Liberation Army.
\end{itemize}
Huawei employees and Chinese diplomats have also received the vaccine, with several other employees of state-owned organizations receiving emergency authorization vaccines~\citep{china5}. 

\subsubsection{Approved Vaccines}
Two vaccines have been approved so far:
\begin{itemize}
    \item BBIBP-CorV (CNBG / Sinopharm) \citep{xia2021safety}
    \item CoronaVac (Sinovac) \citep{zhang2021safety}
\end{itemize}

\subsubsection{Vaccine Administration and Distribution}
Everyone received detailed vaccination information and personalized advice on safety issues before heading to the injection area to receive the long-awaited dose against the coronavirus. Each compartment in the vaccination area is also equipped with a small refrigerator, which is set at 5 degrees to store the vaccine. Each vaccine can be traced back to its source using a code. Data from each vaccination is immediately recorded and then transmitted to China's national system at the Center for Disease Control and Prevention (CDC).

\textbf{Vaccination Rate}
\begin{itemize}
    \item China’s vaccination program started slow but due to the resurgence of COVID-19 cases recently, the process has gained some momentum
    \item China has vaccinated 15.5\% of its people fully, and 43.2\% of its people with at least one dose
    \item 1.1 billion doses have been administered so far
    \item Population to Vaccine ratio: 76.14 COVID-19 vaccine doses administered per 100 people, as dated June 24, 2021 \cite{Ourworldindata}
\end{itemize}

\textbf{Vaccination Governance}: Vaccines are regulated and authorized by the National Medical Products Administration of China~\citep{china10}. 

\subsubsection{Accessibility and Digital Solutions}
 The National Healthcare Security Administration has stated that the COVID-19 vaccine is free to the public. Health insurers and governments cover the cost of the vaccine.
China has established 25,392 vaccination points nationwide, along with detailed regulations and plans for handling and administering the vaccine.
The National Medical Products Administration (NMPA) established a traceability information system to manage approved vaccines.

\subsubsection{Government Strategies and Decisions}
 China joined the COVAX Facility in October~\citep{china9}. 
 National Medical Products Administration (NMPA) and the National Health Commission are in the change to draft laws and regulations for national health policies; These organizations tend to organize the monitoring, evaluation, and handling of adverse drug reactions. The NMPA and provincial drug agencies will implement national and provincial surveillance systems for vaccine traceability.
 
 The National Medical Products Administration (NMPA) and the National Health Commission are in charge of drafting laws and regulations for national health policy; These organizations tend to organize the monitoring, evaluation, and management of adverse drug reactions.
 The NMPA and provincial drug agencies will implement national and provincial surveillance systems for vaccine traceability.

\subsubsection{Money spent}
\begin{itemize}
    \item China has not publicly released its expenditure on the immunization process, thereby rendering any per capita calculations impossible
    \item However, China’s leading vaccine producer Sinopharm has received upwards of \$500 million in funding from private and public investors in the production of the vaccine
\end{itemize}
\cite{china13}

\subsubsection{Variant info}
Though no mutations of the virus have been found to cause significant damage to China, it has reported 18 cases of the double-mutant variant B.1.617 from India. Over resurgence fears, China has increased its lockdown capacity from May 1.
\cite{china14}

\subsection{Turkey}

\subsubsection{Prioritization}
Citizens are mainly being vaccinated according to their risk factors, as determined by the Turkish Science Board. They are mainly following COVAX protocols.
\begin{itemize}
    \item Science board members, healthcare workers, and the elderly over the age of 90 are vaccinated in the first round.
    \item Essential workers in the second round and those who have at least one chronic disease.
    \item Third round - young adults
    \item Fourth round - everyone else
\end{itemize}

\subsubsection{Approved Vaccines}
Two vaccines have been approved so far:
\begin{itemize}
    \item SinoVac \citep{zhang2021safety}
    \item Sinopharm \citep{xia2021safety}
\end{itemize}

\subsubsection{Vaccine Administration and Distribution}
The vaccine is currently used in preventative health care services in Turkey, the highest priority group. The antiserum, the injectors, and the logistic management of the cold chain materials such as transport containers are carried out by the Vaccine Logistics Unit of the Ministry of Health. All vaccines purchased by the Turkish Ministry of Health are evaluated in terms of security before being offered to the citizens, a process that will continue for the COVID-19 vaccines. The evaluation is conducted by Pharmaceuticals and Medical Devices Agency.~\citep{Turkey11}.

The shipping containers are managed by the Department of Vaccines / Antiserum of the Ministry of Health, General Directorate of Public Health for Vaccine-Preventable Diseases.
The vaccines are distributed to each healthcare facility under the control of the Vaccine Tracking System (ATS). The Vaccine Tracking System (ATS) is a national and domestic system in which each dose is identified, and the population and temperature are monitored~\citep{Turkey13}.

\textbf{Vaccination Rate}
\begin{itemize}
    \item Turkey has vaccinated 35.8\% of its population as of 24th of June with at least the first dose and 17.4\% of the people are completely vaccinated
    \item Turkey’s vaccination rate has been unsteady. It started off rapidly but slowed down during the months of March-April in 2021 and has since picked up again
    \item Total doses administered: approximately 44.9 million as of 24th of June 2021
    \item Population to Vaccine ratio: 53.3 COVID-19 vaccine doses administered per 100 people as dated June 24, 2021 \cite{Ourworldindata}
\end{itemize}

\textbf{Vaccination Governance}: Vaccination Governance is administered by the Ministry of Health and Turkish Medicines and Medical Devices Agency.

\subsubsection{Accessibility and Digital Solutions}

The Family Health Center administers the vaccine. It is administered to the general medicine department where the person is registered~\citep{Turkey2}. Each person receives the vaccination application form, from which they can access all the details of the vaccine (electronic vaccination card, etc.) through a text message (SMS) and an e-Nabız account sent to the mobile phone. e-Nabız is an application that enables citizens and health professionals to access health data collected by health care facilities via the internet and mobile devices~\citep{Turkey12}.
Turkey introduced a vaccine tracking system in 2014 to monitor all processes related to vaccines. By using the QR codes required in the country on vaccine doses against any disease, the authorities can monitor storage, delivery, and expiration dates and intervene at any time.
Vaccine Tracking System (ATS) monitors and manages information about which vaccines are given to whom and by whom and monitors potential side effects for immediate intervention.
Adverse reactions following vaccination are recorded electronically in consultation with the nearest public health institution and are subject to careful monitoring and evaluation~\citep{Turkey11}.

\subsubsection{Government Strategies and Decisions}
Turkey has administered 98 percent of all childhood vaccinations needed, giving them a stable vaccination infrastructure. The health ministry has ordered that special vaccination rooms be installed in hospitals and clinics to maintain a safe distance from other medical facilities during the vaccination campaign.

The Ministry of Health has also introduced a mobile application called "aşıla" (vaccination) to track vaccinations. The app, which healthcare workers can use for free, allows them to schedule vaccinations. Through the app, healthcare workers can determine if a candidate is eligible for vaccination and track the schedule for the second dose.

\subsubsection{Money spent}
\begin{itemize}
    \item Turkey has spent \$850mn by ordering and administering the SinoVac vaccine. Turkey’s per-capita expenditure on COVID-19 vaccines so far has been \$10.3 per person
    \item Turkey has also been in talks with Pfizer-BioNTech and plans on ordering an additional 105 million doses of COVID-19 vaccine (though the producer remains undisclosed)
    \item If this producer is to be SinoPharm, Turkey’s per-capita expenditure could triple to approximately \$31/person
\end{itemize}
\cite{Turkey15}

\subsubsection{Variant info}
\begin{itemize}
    \item The UK B.1.1.7 strain, which originated in the United Kingdom, caused an early surge in infections in Turkey
    \item In early April, Turkey's health minister indicated that the B.1.1.7 variation was responsible for 85 percent of all new COVID-19 infections
    \item Many instances in Turkey have recently been linked to the double-mutant variation B.1.617
\end{itemize}
\cite{Turkey16}
\cite{Turkey17}

\subsection{Malaysia}

\subsubsection{Prioritization}
In the vaccine distribution procedure, frontline healthcare workers, the elderly, and individuals with noncommunicable diseases and chronic respiratory disorders would be prioritised. \cite{Malaysia1}.
The distribution of vaccines will be carried out in phases as such \cite{Malaysia12}:

\begin{itemize}
    \item  1st Phase (February – April 2021) 	
Frontline healthcare workers in both public and private healthcare are the first priority category.
Frontline workers in key facilities, defence, and security are the second priority group.
    \item 
 2nd Phase (April – August 2021)
 Priority group 1: Healthcare workers, defence and security personnel who didn't get in Phase-1
 Priority group 2: Senior folks (over 60 years old), those with chronic diseases like heart disease, obesity, diabetes, and high blood pressure, and people with disabilities make up priority group 2.
    \item - 	3rd Phase (May 2021 – February 2022)
 Priority group: Adult population aged above 18 years. 
    
\end{itemize}

\subsubsection{Approved Vaccines}
Malaysia has approved three vaccines \cite{Malaysia2} \cite{Malaysia6} \cite{Malaysia9} \cite{Malaysia10}:
\begin{itemize}
    \item Pfizer-BioNtech
    \item Sinovac
    \item AstraZeneca
\end{itemize}

\subsubsection{Vaccine Administration and Distribution}
Pharmaniaga Bhd, a government-linked company (GLC), is  in charge of the vaccine distribution. From the Kuala Lumpur International Airport (KLIA), the team at Pharmaniaga delivers the vaccines to 5 warehouses in the states: Selangor, Pahang, Penang, Sabah and Sarawak.

Delivery of vaccines from the warehouse to a hospital follows a Just-In-Time (JIT) format according to capacity and vaccination schedules determined by the Malaysian government. JIT is key to avoid stockpiling at hospital storage facilities \cite{Malaysia4}. Pharmaniaga is also prioritizing vaccine supply to the government healthcare institutions before supplying to the private healthcare institutions \cite{Malaysia5}. Currently, the government aims to vaccinate 75,000 people a day across 600 vaccination sites across the country.

\textbf{Vaccination Rate}
\begin{itemize}
    \item 6.3 million total doses administered
    \item 1.73 million fully vaccinated
    \item 5.3\% population fully vaccinated
    \item Population to Vaccine ratio: 19.4 COVID-19 vaccine doses administered per 100 people as dated June 24, 2021 \cite{Ourworldindata}
\end{itemize}

\textbf{Vaccination Governance}: Malaysia's Ministry of Health is in charge of accreditation governance. In Malaysia, the National Pharmaceutical Regulatory Agency (NPRA) is in charge of vaccine testing and registration. On the other hand, the Drug Control Authority (DCA) approves the vaccines' usage based on the NPRA's results. \cite{Malaysia20}

\subsubsection{Accessibility and Digital Solutions}
The COVID-19 vaccines is free for all Malaysians and is not compulsory. Foreigners however will have to pay a charge which is yet to be determined by the Ministry of Health (MOH) \cite{Malaysia13}. Rural residents are more likely to receive vaccines from other licensed vaccines that do not require severe cold storage, such as the Pfizer-BioNtech vaccine and the AstraZeneca vaccine, which must be stored between 2°C and 8°C.

The contact tracing app known as MySejahtera is used to coordinate vaccine administration. The app plays a key role in registering people who are eligible to receive the COVID-19 vaccine and is used to monitor them post-vaccination for adverse events as there are a high number of users on the app \cite{Malaysia14}. The app has a sign-up feature to get the user's consent as well as register the user for vaccination. Reporting of adverse events is done on a self-reporting basis through the app \cite{Malaysia16}.

Those without access to smartphones will receive a card upon their first vaccination, which must be kept. This card will serve as a record of vaccination and remind them to come for their second dose on the date specified on the card itself. \cite{Malaysia15}

It is currently unclear how the data is stored but all personal data collected by the app complies with current laws \cite{Malaysia14}.

\subsubsection{Government Strategies and Decisions}
The Malaysian government has set aside US\$504 million to purchase enough vaccines to protect 26.5 million citizens (about 80\% percent of total population). The Malaysian government aims for 80\% of its population to be vaccinated. This would mean that some portion of children under 16 years old would have to be vaccinated. However, they are still waiting for Phase 3 trials to be completed for this specific age group \cite{Malaysia14}. On top of that, vaccinations will not be made mandatory. The Malaysian government has repeatedly emphasized that vaccinations will follow a voluntary basis \cite{Malaysia17}. 

\subsubsection{Money spent}
\begin{itemize}
    \item The government tapped into the national trust fund to help fund its vaccine purchases. The fund totalled \$4.6 billion, of which national oil company Petronas contributed \$2.5 billion
    \item The government has raised vaccine allocation to \$840 million from \$550 million and expects to spend an additional \$360 million on the immunization program -- more than double the initial forecast
\end{itemize}
\cite{Malaysia21}

\subsubsection{Variant info}
The highly infectious COVID-19 variant B.1.617 that originated from India was detected in Malaysia as a first in early May during a screening of Indian nationals at the Kuala Lumpur International Airport.
\cite{Malaysia22}

\subsection{Germany}

\subsubsection{Prioritization}
Germany has established three priority groups for vaccination:

\begin{itemize}
    \item Group 1 (Highest Priority): over 80 years old, or elderly-care workers, or healthcare workers with any relation to COVID-19 
[5.4 M People of this group got their first dose of vaccination. Vaccination for this high priority group is almost completed] (March-9) \cite{Gm0}
    \item Group 2 (Higher Priority): over 70 years old, or deadly underlying health conditions, or live/in contact with pregnant women, or doctors and other healthcare workers, police, teachers and essential workers who maintain hospital infrastructure (Vaccination for this group is going on currently)
    \item Group 3 (High Priority): over 60 years old, or underlying health conditions like diabetes, asthma, or all other healthcare workers and essential workers who are not there in group-2 (Vaccination for this group can be started around May)
    \item Group 4 Everyone else, who did not get vaccinated. [Will be started in July and probably completed by the end of September]
\end{itemize}

\subsubsection{Approved Vaccines}

\begin{itemize}
    \item Moderna
    \item Oxford/AstraZeneca
    \item Pfizer/BioNTech
\end{itemize}

\subsubsection{Vaccine Administration and Distribution}
Vaccines are delivered to a central location within Germany and thereafter distributed to the federal states with respect to population sizes. There are 27 designated delivery points that the government has isolated.
Germany has sent "mobile" vaccination teams to elderly care homes with many vaccination centers closed right now. 
[1] Load boxes with each vaccine to defrost overnight
[2] Deliver vaccines (84 shots in each box) within a few hours or else it will be useless
[3] Repeat

\textbf{Vaccination Rate}
\begin{itemize}
    \item 69.5 million total vaccine doses administered
    \item 27.8 million (around 33.2\% of population) fully vaccinated
    \item Population to Vaccine ratio: 82.9 COVID-19 vaccine doses administered per 100 people as dated June 24, 2021 \cite{Ourworldindata}
\end{itemize}

\textbf{Vaccination Governance}: 
The European Medicines Agency (EMA) previously issued a corresponding recommendation for the vaccines, which the European Commission approves. The vaccine batches are delivered to a central location in Germany and subsequently released by the Paul Ehrlich Institute.

\subsubsection{Accessibility and Digital Solutions}
With € 2.7 billion set aside for vaccination, Germany has made the coronavirus vaccine free for all individuals residing in Germany (either through health insurance or through German registration certificate). At the time of writing, only people within priority groups (see "Priority") can request for vaccination. 

Most states offer two options: online appointment or appointment by phone. In Berlin, only people who have received a letter of invitation from the Senate Health Administration can request an appointment.

The Robert Koch Institute (a Germany federal government agency) has released an app called Corona-Warn App [18] that utilizes the GAEN 2.0 framework to implement contact-tracing within their country.

\subsubsection{Government Strategies and Decisions}
The current motto of the German government is "Vaccinate-Vaccinate-Vaccinate." Though vaccination is not mandated, this push is at the crux of many government decisions in Germany.

\subsubsection{Money spent}
\begin{itemize}
    \item The German government intends to spend almost 9 billion euros (\$10.9 billion) this year alone to assist in the procurement of up to 635 million COVID-19 vaccine injections for its citizens and other EU member states
    \item Bettina Hagedorn, the deputy of Finance Minister Olaf Scholz, has made a proposition to the lawmakers in a letter demanding that they approve a request of an additional 6.22 billion euros by the Health Minister
    \item Jens Spahn for acquiring more doses. This is an addition to the pre-existing 2.66 billion euros that have already been earmarked for the immunization programme in the 2021 budget
\end{itemize}
\cite{Gm14}

\subsubsection{Variant info}
\begin{itemize}
    \item Cases of B.1.617 variant of COVID-19, which was first found in India
\end{itemize}
\cite{Gm15}

\subsection{Singapore}

\subsubsection{Prioritization}

The Singaporean government has decided to prioritize the following groups of people in their vaccination program(In order of highest to lowest priority)\cite{Singapore5}:

\begin{itemize}
    \item Healthcare frontliners
    \item Elderly above 70 years old
    \item Singaporeans with vascular medical comorbidities
    \item Elderly from 60 - 69 years old
    \item Personnel in essential services (i.e water, utilities, etc.)
    \item Other Singaporeans under 60 years old and long-term residents

\end{itemize}

\subsubsection{Approved Vaccines}
 Available Vaccines in Singapore: 

\begin{itemize}
    \item Arcturus, Pfizer, Moderna: mRNA type
    \item Astra-Zeneca, Gamaleya, Can-Sino: Virus Vectored
    \item Sinovac, Bharat-BioTech: Inactivated Virus
    \item Novavax, Sanofi-Pasteur: Protein Subunit Vaccines
\end{itemize}

\subsubsection{Vaccine Administration and Distribution}

It has been estimated that they should have enough vaccine supply for their whole population by the third quarter. The Singaporean Government has announced that all Singaporeans and long-term residents in Singapore will be vaccinated by the end of 2021 for free. Prior-booking is required for Singaporeans to get vaccinated when it’s their turn and is necessary due to the cold-chain requirements

\textbf{Vaccination Rate}
\begin{itemize}
    \item 5 million doses have been administered in Singapore so far
    \item Almost 35.1\% of the country is fully vaccinated
    \item People of Singapore are getting vaccinated at a rate of 45,636 doses per day \cite{Singapore14}
    \item Population to Vaccine ratio: 85.96 COVID-19 vaccine doses administered per 100 people as recorded on June 24, 2021. \cite{Ourworldindata}
\end{itemize}

\textbf{Vaccination Governance}: 
Before authorising the vaccine, the Health Sciences Authority (HSA) and the Expert Committee on COVID-19 Vaccination scrutinised the clinical evidence.
Under the Pandemic Special Access Route, the Health Sciences Authority (HSA) approved an interim licence for the COVID-19 vaccinations to be used in Singapore for the prevention of Coronavirus Disease 2019 (COVID-19) in people aged 18 and up (PSAR).

\subsubsection{Accessibility and Digital Solutions}
Singapore is setting up special centers where large numbers of people are vaccinated against COVID-19 every day.  The delivery is easily accessible to all citizens at any time following the priority hierarchy. The Ministry of Health will send a personalized letter notifying you that it’s your turn to be vaccinated \cite{Singapore9}.

Singaporeans can either choose to book online at the vaccine.gov.sg website or book physically at a local community centre .The Singaporean government is concurrently preparing clinics and vaccination centres for COVID-19 vaccination. 

When patients arrive for vaccinations they are given a vaccine information sheet and vaccine selection form. Before their turn, they are asked to fill out the first part of the Vaccination Screening Form for personal information, medical information, declaration and consent.

Once they've received the first shot, they are given a vaccination card \cite{Singapore8}. These vaccination cards are given to patients after the first dose. The card serves to provide information such as:

Type of vaccine administered
\begin{itemize}
    \item Appointment dates for the second dose
    \item Brief post-vaccination advice
\end{itemize}

Singaporeans that choose to get inoculated also have the option to check vaccination records online. These vaccine records will be updated in Singapore’s National Immunization Registry. The Singaporean government mainly relies on the Health Science Authority (HSA) to report adverse events post-vaccination. The Vaccine Adverse Event (VAE) Online. Reporting Platform seems to be the main digital platform to report adverse events post-vaccinations \cite{Singapore11}.

\subsubsection{Government Strategies and Decisions}
Singapore has set aside roughly S\$1 billion (US\$754 million) for vaccines[1].They’ve also publicly announced that Singaporeans will not be given the choice to pick which COVID-19 vaccines they would like as this further complicates the process. Incase of any injury during the vaccination process, the victim will get Financial Assistantship. Singaporeans will take the vaccines on a voluntary basis \cite{Singapore7}.

\subsubsection{Money spent}
\begin{itemize}
    \item Singapore Government announced a 8.3 billion dollar package to support the Covid Vaccination efforts
\end{itemize}
\cite{Singapore12}

\subsubsection{Variant info}
\begin{itemize}
    \item Singapore witnessed the B.1.1.7 variant, which was initially found in UK during December of 2020
    \item B.1.617 variant from India has been detected in Singapore recently which lead to the shutting down of schools
\end{itemize}
\cite{Singapore13}

\subsection{Hong Kong}

\subsubsection{Prioritization}

\begin{itemize}
    \item Residents and employees in nursing homes for the elderly and people with disabilities, as well as other institutional services Staff in health-care settings
    \item workers in other critical healthcare services that are at a higher risk of COVID-19 exposure and people aged 60 and above
    \item People between the ages of 16 and 59 who have chronic medical conditions
\end{itemize}
\cite{HK4}

\subsubsection{Approved Vaccines}

\begin{itemize}
    \item CoronaVac (by Sinovac Biotech)
    \item Tozinameran (by BioNTech-Fosun Pharma)
    \item ChAdOx1 nCoV-19 (by AstraZeneca-Oxford)
\end{itemize}

\subsubsection{Vaccine Administration and Distribution}

The program's goal is to provide vaccines to approximately 7.5 million Hong Kong residents by 2021. The public has the choice of being vaccinated or not being vaccinated. The vaccines will arrive in Hong Kong in batches and be distributed to private clinics and hospitals.

\textbf{Vaccination Rate}
\begin{itemize}
    \item It is estimated that, people of Hong Kong are getting vaccinated at a rate of 29,515/day
    \item Till date, around 1.32 million people have been fully vaccinated in Hong Kong, which approximates to 17\% of the country’s population \cite{HK9}
    \item Population to Vaccine ratio: 44.66 COVID-19 vaccine doses administered per 100 people, June 24, 2021 \cite{Ourworldindata}
\end{itemize}

\textbf{Vaccination Governance}: 
The Government of Hong Kong has set up the Advisory Panel on COVID-19 Vaccines, with members appointed by the Chief Executive. The Advisory Panel provides expert advice on recommendation of authorization of the vaccines to the Government, which is made available for public information.

\subsubsection{Accessibility and Digital Solutions}
Private clinics and hospitals are keeping digital records to be shared with the government, although it may not be available to private medical institutions. Hong Kong has not yet provided specifics on this, however.
The government will introduce a free COVID-19 vaccination programme for all Hong Kong citizens throughout the territory. COVID-19 vaccines will be distributed via the programme in a variety of environments, including hospitals, clinics, outreach to institutions, and neighbourhood vaccination centres. The Department of Health (DH) has an adverse drug reaction (ADR) reporting system which receives adverse events following immunization (AEFIs) reports to monitor the safety of COVID-19 vaccines.COVID-19 Electronic Vaccination and Testing Record System. This system facilitates free download by the general public of their own electronic testing records for the COVID-19 testing conducted within 31 days. The system will add a function to allow checking of electronic vaccination records. By that time, the public can download both of their COVID-19 electronic vaccination records and testing records after verifying their identity through “iAM Smart”. The system will not store or check the public’s other vaccination records.
One can download the electronic testing record by inputting the personal information (such as Hong Kong identity card number, specimen bottle number or laboratory number/specimen number, etc.) of one’s into the system.

\subsubsection{Government Strategies and Decisions}
During vaccination campaigns, the number of measures that are used. To tie up with the vaccination programme and document any adverse events that occur to the patient associated with the relevant vaccine delivery, the government established the Expert Committee on Clinical Events Assessment Following COVID-19 Immunisation. The Expert Committee will monitor any potential adverse events that occur after COVID-19 vaccines are administered and provide clinical advice and recommendations on how to assure the safety of the licenced vaccines [6]. The Hong Kong government's ultimate goal is to distribute the COVID-19 vaccine to the majority of Hong Kong's population by 2021.  \cite{HK6}

\subsubsection{Money spent}
\begin{itemize}
    \item The Hong Kong government allocated around 8.4 billion dollars for Covid-19 Vaccination in their budget
\end{itemize}
\cite{HK7}

\subsubsection{Variant info}
\begin{itemize}
    \item The mutated B.1.617 variant originated in India was found in Hong Kong in April, 2021
\end{itemize}
\cite{HK8}

\subsection{Italy}

\subsubsection{Prioritization}

Italy will prioritize the following population groups:

\begin{itemize}
    \item Health and social care workers. 
    \item Elderly above 70 years old.
    \item Residents and staff of residential care facilities for the elderly. 
    \item Older persons.

\end{itemize}

\subsubsection{Approved Vaccines}
 Approved Vaccines in Italy are: 

\begin{itemize}
    \item Moderna
    \item Oxford/AstraZeneca
    \item Pfizer/BioNTech
    \item Johnson and Johnson
\end{itemize}

\subsubsection{Vaccine Administration and Distribution}

A distribution model focused on a single national storage site and a collection of second level territorial sites will be used for vaccines that need a normal cold chain (2°/8°C). Vaccines that require an extremely cold chain (-20°/-70°C) would instead be supplied directly by the manufacturer to the territory's vaccination points (approximately 300 at the moment), in accordance with Regions.The initial phase of the campaign involves a centralized management of vaccination, using hospitals and mobile units to vaccinate those who cannot independently reach the centers set up for vaccination. The administration plans to involve territorial clinics, general practitioners, pediatricians , military health department and doctors working for private companies \cite{It3}, in addition to mobile units \cite{It6}.
The vaccine is to be administered in two doses, with the supply sufficient for 101 million people in total. Vaccines are not purchased by states, but by the European Commission, and divided according to population size. Italy is entitled to 13.5\% of the doses bought by the EU \cite{It2}

\textbf{Vaccination Rate}
\begin{itemize}
    \item Around 47.7 million doses have been administered in Italy
    \item 16.6 million people (or 27\% of the population) are fully vaccinated
    \item The rate of vaccination in Italy is 497,487 doses per day \cite{It12}
    \item Population to Vaccine ratio: 78.93 COVID-19 vaccine doses administered per 100 people, as of June 24, 2021. \cite{Ourworldindata}
\end{itemize}

\textbf{Vaccination Governance}: 
A strategy plan for anti-SARS-CoV-2 / COVID-19 vaccination was prepared by the Ministry of Health, the Extraordinary Commissioner for Emergency, the National Institute of Health (ISS), the National Agency for Regional Health Services (AGENAS), and the Italian Medicines Agency. (AIFA).

\subsubsection{Accessibility and Digital Solutions}
Italy did not set up a national information system to make reservations. Instead, each of the 20 regions built its own booking platform \cite{It4}. Only the Italian Postal service developed a platform to manage bookings and Sicily is the first region using it.\cite{It5} To track vaccinations, the Italian Minister of Public Health Official is using a dashboard powered by Microsoft PowerBi.

\subsubsection{Government Strategies and Decisions}
On December 2, 2020, the Italian Ministry of Health, the Covid-19 Emergency Commissioner, and the Italian Medicines Agency presented the Strategic Plan for SARS-CoV-2/COVID-19 vaccine protocols to the Italian Parliament.
Here are the key points of the Plan:
\begin{itemize}
    \item Mfree and guaranteed vaccination for all: December 27, 2020, start of vaccination in Italy and Europe (Vaccine Day)
    \item identification of the categories to be vaccinated with priority in the initial phase of limited availability of vaccines: health and social workers, residents and staff of RSAs for the elderly
    \item The Special Commissioner is in charge of item logistics, procurement, storage, and transportation
    \item governance of the immunisation plan will be ensured by regular cooperation between the Ministry of Health, the Emergency Commissioner, and the Regions.
    \item Setting up an information system to manage the vaccination campaign in an effective, integrated, secure and transparent way
    \item pharmacosurveillance and immunological surveillance to ensure the highest level of safety throughout the vaccination campaign and the immune response to the vaccine. \cite{It1}
\end{itemize}

AIFA (Italian Medicines Agency) will promote the launch of some independent post-authorization studies on COVID-19 vaccines. Vigilance activities will concern both the collection and evaluation of spontaneous reports of suspected adverse reactions (passive pharmacovigilance) and proactive actions, through pharmaco-epidemiology studies/projects (active pharmacovigilance). AIFA has set up a Scientific Committee, which, for the entire period of the vaccination campaign, will have the function of supporting the Agency in the phase of setting up activities, in the overall analysis of the data that will be collected and in the identification of possible interventions. The aim is to have, also through an international collaborative network, the ability to highlight any sign of risk and, at the same time, to compare the safety profiles of different vaccines that will become available, to provide recommendations. \cite{It8}

\subsubsection{Money spent}
\begin{itemize}
    \item 400 million euros will be spent on vaccinations and medications to treat COVID-19 patients, while 70 million euros will be spent on rapid tests
    \item Italy is slated to get the largest share of the European Union's recovery fund, with a portion of these funds going toward addressing some of the long-standing challenges that are largely perceived as holding Italy and its economy behind
\end{itemize}
\cite{It11}

\subsubsection{Variant info}
\begin{itemize}
    \item Around September 2020, UK’s B.1.1.7 variant was found in Italy \cite{It9}
    \item Indian variant has been detected in the northern parts of Italy in April 2021 \cite{It10}
\end{itemize}

\subsection{Vietnam}

\subsubsection{Prioritization}
The 11 priority groups include:
\begin{itemize}
    \item health professionals
    \item pandemic prevention and fight task force (members of steering committees, people
working in quarantining areas, reporters and others)
    \item diplomats, and customs and immigration staffs, military personnel, police force,
teachers
    \item elderly aged over 65
    \item those working to maintain the supply of essential services such as aviation,
transportation, tourism, electricity and water supply
    \item people with chronic diseases
    \item people going abroad on a mission, to work, or study
    \item those in pandemic-hit areas

\end{itemize}

\subsubsection{Approved Vaccines}
 Approved Vaccines in Vietname are:

\begin{itemize}
    \item Sputnik V
    \item Oxford/AstraZeneca
\end{itemize}

\subsubsection{Vaccine Administration and Distribution}
Around 600,000 Vietnamese people, comprising healthcare and COVID-19 combat workers, will be given shots in the first quarter. Meanwhile, 1.8 million customs officers, diplomats, army and police personnel, and teachers will be injected in the second quarter. During the
second half, over 16 million teachers and people above 65 will access 33 million doses of vaccines to be provided by the COVAX Facility later. The government plans free vaccinations. \cite{VT1}
Hospitals and health centers at the central, provincial, city, district levels administer injections. At the grassroots level, commune health stations are organizing a campaign to vaccinate COVID-19 through mobile stations and injection sites. \cite{VT2}

\textbf{Vaccination Rate}
\begin{itemize}
    \item Only 1.5\% (approximately 3 million) of Vietnam’s population is fully vaccinated as of now at a speed of approximately 6000 doses per day  \cite{VT9}
    \item Population to Vaccine ratio: 3.00 COVID-19 vaccine doses administered per 100 people, as of June 24, 2021 \cite{Ourworldindata}
\end{itemize}

\textbf{Vaccination Governance}: 
The Centers for Disease Control in provinces and cities synthesize the list of immunization units and the number of vaccinated subjects in the provinces and cities and notify the Department of Preventive Medicine, the Central Institute of Hygiene and Epidemiology to be supplied with vaccines. Implementation time is March 7-10, 2021.
Based on the plan of the Ministry of Health, the provincial health departments of the provinces and cities take the lead in developing the local COVID-19 vaccine use plan and directing the preparation of the list of injecting subjects according to risk groups within 7 days after the Ministry of Health issued the Plan on receipt, transportation, storage and use of COVID-19 vaccine. \cite{VT2}

\subsubsection{Accessibility and Digital Solutions}
The Ministry of Health has set up a webpage specifically dedicated to COVID-19 influenza disease section. WHO also offers a curation of situation reports. As of April 6, 2021, there are up to Report 36.
The Ministry of Health, in collaboration with the Ministry of Information and Communications, has also launched a mobile application called NCOVI. According to its Google Play page, the app seeks to “assist people nationwide to make voluntary medical declarations, contributing to the prevention and fight against acute pneumonia caused by new strains of COVID-19.” The app will also serve as an official communication channel for state agencies. \cite{VT6}

\subsubsection{Government Strategies and Decisions}
On the evening of April 7, the Ministry of Health issued Decision No. 1821 / QD-BYT on the distribution of COVID-19 round 2. Accordingly, there are 811,200 doses of Astra Zeneca's vaccine funded by COVAX. \cite{VT3}
The Prime Minister decided to supplement the state budget estimate in 2021 to the Ministry of Health 502.9 billion VND from the central budget reserve in 2021 to continue the procurement of materials, chemicals, bio-products and equipment. medical equipment
necessary for the prevention and control of COVID-19 epidemics according to the plan approved by the National Steering Committee for COVID-19 epidemic control in 2020. The Ministry of Health is fully responsible for the management, use and settlement of the
additional funding; Carry out procurement in accordance with the provisions of the law, for the right purposes and subjects, to ensure thrift, efficiency, publicity, transparency and in
accordance with the epidemic's manifestation in 2021. \cite{VT4}

\subsubsection{Money spent}
\begin{itemize}
    \item Vietnam government spend more than 1 Billion dollars on vaccination to get 150 Million doses
\end{itemize}
\cite{VT7}

\subsubsection{Variant info}
\begin{itemize}
    \item Vietnam has witnessed B.1.222, B.1.619, D614G, B.1.17, B.1.351, A.23.1 and B.1.617.2 which includes both Indian and UK variant
    \item Recently a new variant which seems like a hybrid of Indian and UK mutants
\end{itemize}
\cite{VT8}

\subsection{Thailand}

\subsubsection{Prioritization}

\begin{itemize}
    \item Front-line health workers
    \item  Patients with acute and chronic respiratory disorders, chronic obstructive lung disease, and difficult-to-control asthma are all included in this category. 
    \item patients with cardiovascular disorders, chronic kidney disease at stage 5 or higher, stroke patients, and cancer patients undergoing chemotherapy
    \item persons suffering from diabetes or obesity
    \item staff involved in disease prevention and control Exposure to COVID-19

\end{itemize}
\cite{TH1}

\subsubsection{Approved Vaccines}
 Approved Vaccines in  

\begin{itemize}
    \item Coronavac
    \item Oxford/AstraZeneca
    \item Pfizer/BioNTech
    \item Johnson and Johnson
\end{itemize}

\subsubsection{Vaccine Administration and Distribution}
There are two phases of vaccination. The first phase's target groups include frontline medical and public health workers, chronic disease patients, people over the age of 60, and disease control authorities. Non-frontline medical and public health professionals, tourism industry employees, international travellers, the general public, ambassadors, representatives of international organisations, foreign business people, long-stay visitors, and manufacturing and service sector workers are among the target categories in the second phase. The Cabinet also authorised the Ministry of Public Health's plan to increase its order of AstraZeneca vaccine from 26 million to 61 million doses. \cite{TH2}
Thailand has also authorised the private sector to import vaccines to complement the national vaccination programme. Despite criticisms and concerns regarding black market formation and inequitable access to vaccines, the government holds a lax attitude towards such demand from the private sector
“as it would help accelerate the progress of our inoculation program.” \cite{TH4} Thailand's private hospitals and drugmakers are awaiting government approval to manufacture some COVID-19 vaccines on their own to satisfy demand from affluent Thais and businesses who want their shots as soon as possible. Just two vaccines, made by China's Sinovac and the United Kingdom's AstraZeneca, will be administered free of charge via public hospital networks under the government's vaccination policy, which began March 1. For those who are able to pay, private companies are now offering expedited vaccinations. The government is being accommodating when it comes to vaccine imports.
So far, at least one pharmaceutical firm based in the United States and a hospital operator have applied for import licences. \cite{TH4}

\textbf{Vaccination Rate}
\begin{itemize}
    \item 2.45 Million people (approximately 3.51 \%) of Thailand’s population is fully vaccinated as of now at a speed of approximately 98713 doses per day. \cite{TH10}
    \item Population to Vaccine ratio: 12.40 COVID-19 vaccine doses administered per 100 people, as of June 24, 2021. \cite{Ourworldindata}
\end{itemize}

\textbf{Vaccination Governance}: 
The Center reports vaccination progress for COVID-19 Situation Administration (CCSA). The Ministry of Public Health oversees vaccine distribution. It expects to distribute 800,000 doses of Sinovac's COVID-19 vaccine across the country by April 1, 2021. The first 22 provinces will begin immunisation on April 1st, followed by the rest of the provinces in June. \cite{TH1}

\subsubsection{Accessibility and Digital Solutions}
The Ministry of Health plans to issue “vaccine passports” to residents who have completed their COVID-19 vaccinations so that they can travel abroad. For vaccinated visitors from abroad, Thailand also plans to shorten the mandatory quarantine period from 14 days to 7 days. Starting October 2021, vaccine passport holders may be completely waived from quarantine upon arrival. For travelers without vaccine passports but with negative test results, they still need to quarantine for 10 days upon arrival.

In May 2021, the Ministry of Public Health will launch "Mor Prom (Health Wallet)" a mobile application that allows citizens to make COVID-19 vaccination appointments. The app, expected to launch on May 1, allows users to book their free jabs at designated state hospitals and public health stations, from June. It also includes a feature that enables post-vaccination monitoring to check for side effects. A recent survey found that 80\% of people living in Bangkok use phone apps as their main means of communication, a number that falls to 50\% for people outside the capital. \cite{TH6}

\subsubsection{Government Strategies and Decisions}
The Thai cabinet agreed on Sunday not to join the WHO-sponsored coronavirus vaccine scheme, claiming that doing so would result in the nation spending more on vaccines and putting delivery times in jeopardy. Responding to media claims that Thailand is the only Southeast Asian nation to miss the WHO's COVAX system, government spokesman Anucha Buraphachaisri said that Thailand is not eligible for free or low-cost vaccinations as a middle-income country under the initiative. \cite{TH3}

\subsubsection{Money spent}
\begin{itemize}
    \item The government of Thailand announced a budget of around 1.9 billion dollars to support the COVID-19 vaccination in Thailand
\end{itemize}
\cite{TH7}

\subsubsection{Variant info}
\begin{itemize}
    \item UK variant has been found in Thailand in April 2021
\end{itemize}
\cite{TH8}
\cite{TH9}

\subsection{Norway}

\subsubsection{Prioritization}
Norway’s vaccine prioritization guidelines are as follows:
\begin{itemize}
    \item  Nursing home residents and selected groups of healthcare personnel
    \item Age $\geq$ 85 years as well as certain types of healthcare workers, are at risk.
    \item Age 75 to 84 years
    \item People between the ages of 65 and 74, as well as people between the ages of 18 and 64 who have major diseases
    \item Age 55 to 64 years
    \item Age 45 to 54 years
    \item Age 18-44 years
\end{itemize}

\subsubsection{Approved Vaccines}
 Approved Vaccines in Norway are \cite{Nw1}: 

\begin{itemize}
    \item Moderna
    \item Oxford/AstraZeneca $\ast$
    \item Pfizer/BioNTech
    \item Johnson and Johnson
\end{itemize}
Authorities halted the introduction of the AstraZeneca vaccination on March 11, 2021, after a small group of younger people were hospitalised due to problems such as blood clots, haemorrhage, and low platelet counts. \cite{Nw4} The Norwegian Institute of Public Health (FHI) advised a permanent suspension of the AstraZeneca vaccination in Norway on April 15 due to a rare but serious risk of problems with the vaccine. The Norwegian government is still deliberating over the recommendation, given the WHO advice to continue administering the vaccine. \cite{Nw7}

\subsubsection{Vaccine Administration and Distribution}
The dispatch, packaging and distribution of the vaccine doses must be planned well in advance, as the vaccines currently planned for usage in Norway require sub-zero storage and transportation. Norway Institute of Public Health (NIPH) or Folkehelseinstituttet (FHI) is based on a distribution key that includes the geographical distribution of the population in priority groups, based on their share of the total population. The model aggregates population density, hospital admissions, and risk group proportions. However, while it does account for distribution of risk groups per municipality, it does not take into account the geographical distribution of health personnel. \cite{Nw2}

\textbf{Vaccination Rate}
\begin{itemize}
    \item A total of around 3.9 million doses have been administered in Norway. Vaccination is happening at a rate of 38,281 doses per day \cite{Nw10}
    \item Population to Vaccine ratio: 72 COVID-19 vaccine doses administered per 100 people, as of June 24, 2021. \cite{Ourworldindata}
\end{itemize}

\textbf{Vaccination Governance}: 
The Norwegian Institute of Public Health (FHI) is responsible for publishing coronavirus guidelines, vaccine information, vaccine approvals, and conducting vaccination priority and distribution guidelines. However, the Norwegian government has final say in official policies regarding vaccination roll-out and guidelines and also considers WHO guidelines. \cite{Nw5}

\subsubsection{Accessibility and Digital Solutions}
A downloadable vaccination card will be available on the vaccine service provided by helsenorge.no. The vaccination stage will be visible on the card. If an individual is not registered with helsenorge.no, a simplified vaccination card is provided by the vaccinator where the vaccinated individual must remember when their second dose takes place. The service is available in English and Norwegian. \cite{Nw3}

The EU plans to implement vaccine certificates, both digital and physical, to travel across all EU borders and further is appealing to the Norwegian government to opt into the EU plan. The certificate will utilize an encrypted QR code and is collaborating with the WHO to ensure international recognition of the EU certificate. \cite{Nw6}

\subsubsection{Government Strategies and Decisions}
The Norwegian government oversees the implementation of any policies and guidelines regarding coronavirus vaccination. The Norwegian Institute of Public Health (FHI) is responsible for researching and providing recommendations about vaccine guidelines and decisions. \cite{Nw5} However, the Norwegian government takes into account multiple factors, such as WHO guidelines, before implementing FHI recommendations. \cite{Nw7}

\subsubsection{Money spent}
\begin{itemize}
    \item Norway has donated 2.2 billion crowns to the Oslo-based Coalition for Epidemic Preparedness Innovation (CEPI), which aims to prevent epidemics by funding vaccine research. 
    \item Norway will contribute \$1 billion to the global rollout of any COVID-19 vaccine, as well as vaccinations for other diseases.
    \item GAVI, the Global Partnership for Vaccines and Immunization, a global alliance of commercial and governmental organisations dedicated to global immunisation, will receive \$1 billion in direct support from Norway between 2021 and 2030.
    \item Norway has given the World Health Organization an additional 50 million crowns (\$4.8 million) to fight the novel coronavirus outbreak
\end{itemize}
\cite{Nw9}

\subsubsection{Variant info}
\begin{itemize}
    \item A new variant was found in Norway in February, 2021 called the B1525 variant
    \item The B1525 has close resemblance with B117 and B1351 variants found in UK and South Africa respectively
\end{itemize}
\cite{Nw8}

\subsection{Russia}

\subsubsection{Prioritization}
Priority vaccination against COVID-19 applies to the following groups: 
Employees of medical institutions (all), educational institutions, the police, local public transport, retail, charities, public catering companies and other organizations whose work involves direct contact with a large number of people includes (hotels, hairdressers, dry cleaners, banks, security companies and others). \cite{Russ1}

The age of the vaccinated is limited to 60 years. People with underlying health problems, pregnant women, people with respiratory diseases in the last two weeks are excluded from vaccination. \cite{Russ2}

\subsubsection{Approved Vaccines}
 Approved Vaccines in  

\begin{itemize}
    \item Sputnik V
    \item EpiVacCorona
    \item Pfizer/BioNTech
    \item CoviVac
\end{itemize}

\subsubsection{Vaccine Administration and Distribution}
The Russian-made Sputnik V vaccine has been distributed at 70 clinics around Moscow with the Kremlin stating that mass vaccinations will be voluntary and free of charge, beginning December 2022. The vaccine in its lyophilized dry form can be stored at temperatures being +2 and +8 degrees celsius. The cost for international markets will be \$10 per dose. \cite{Russ2}

Based to the Russian Direct Investment Fund (RDIF), Russia was the most trusted vaccine production country (54 percent) as of March 24, 2021, according to the results of a YouGov survey of 9,417 respondents from 9 countries. Furthermore, the survey indicated that 33.2\% of respondents would prefer the Sputnik V vaccine, second only to Pfizer/BioNTech at 36.6\%. \cite{Russ4}

Due to a vaccine surplus in Moscow, the government is offering appointment-free walkins to get the Sputnik V vaccine free of charge. They are further incentivizing it with a free ice pop upon vaccination. [9] Furthermore, in order to reach limited-mobility residents, Russia is offering at home vaccination visits beginning April 13.  \cite{Russ3}

Putin announced that in early March, all but 9 Russian regions have begun to deploy the vaccine. The delay has been attributed to logistical concerns regarding the locations of the regions. \cite{Russ7}

\textbf{Vaccination Rate}
\begin{itemize}
    \item As of June 24, 2021, 11\% (16.29 million) of Russia’s population is fully vaccinated at a speed of 312,634 doses per day \cite{Russ13}
    \item Population to Vaccine ratio: 25.33 COVID-19 vaccine doses administered per 100 people, 24 June, 2021 \cite{Ourworldindata}
\end{itemize}

\textbf{Vaccination Governance}: 
The Russian Health Ministry oversees and recommends health decisions and advice regarding covid guidelines, treatment, and vaccination rollout.

\subsubsection{Accessibility and Digital Solutions}
Although not many details about the country’s accessibility and digital initiatives have been made public, in terms of vaccine passports, Russia’s foreign minister has voiced concerns about such passports infringing upon people’s right to travel. Russia is currently issuing proof of vaccination documents. \cite{Russ8}

\subsubsection{Government Strategies and Decisions}
The Russian government has approved three vaccines for use, by recommendation of the Russian Health Ministry. Furthermore, the Health Ministry provides input on vaccination priority group phase openings. \cite{Russ5}

\subsubsection{Money spent}
\begin{itemize}
    \item The Russian government reported in July 2020 that it has spent over 116 billion rubles (\$1.67 billion) battling the pandemic \cite{Russ9}:
    \begin{itemize}
        \item 76 billion rubles were invested across the country to equip beds for COVID-19 patients
        \item Free delivery of test systems to the government, municipal medical institutions, and treatment of coronavirus-infected patients were allotted 40 billion rubles.
    \end{itemize}
    \item The Russian Direct Investment Fund (RDIF), the organisation in charge of global Sputnik V vaccine manufacturing and supply partnerships, announced in April 2021 that the country is already allowing free technology transfer and capital expenditure support to its Covid-19 vaccine manufacturing partners in other countries. Twenty vaccine companies from ten nations, including Korea, China, and India, are members of Russia's vaccine collaboration
    \cite{Russ10}
\end{itemize}

\subsubsection{Variant info}
\begin{itemize}
    \item The Russian government has not officially recorded confirmed cases of variants. Currently there are two pieces of conflicting reports
    \item The Kommersant newspaper reported 16 cases of the variant among Indian students at Ulyanovsk State University, some 700 kilometres (435 miles) east of Moscow. Tatiana Golikova, the country's deputy prime minister, had stated that no cases of the variation had been reported \cite{Russ11}
    \item However, the regional branch of consumer health watchdog Rospotrebnadzo on Thursday denied the Kommersant report \cite{Russ12}
\end{itemize}

\subsection{Egypt}

\subsubsection{Prioritization}
Egypt started vaccinating its medical staff and front-line workers. The vaccine was then made available to elderly people and patients with chronic diseases by early March 2021 \cite{Eg1}.
Egypt plans on vaccinating close to 65\% of its population under the age of 35, and 40 million people in the 35-40 year range by the end of 2021.Immunocompromised patients, children and pregnant women will not be administered vaccines in the initial phases of the program before the side-effects of the vaccine are studied in greater detail \cite{Eg2}.

\subsubsection{Approved Vaccines}
 Approved Vaccines in  

\begin{itemize}
    \item China/Sinopharm
    \item Oxford/AstraZeneca
    \item Pfizer/BioNTech
    \item Russia/Sputnik IV
\end{itemize}

\subsubsection{Vaccine Administration and Distribution}
Egypt had earlier set up 34 distribution centers across the country, but as the demand increases, it continues to set up more facilities. It currently has 139 centers but the government plans on increasing that to 339 by May \cite{Eg5}. 

\textbf{Vaccination Rate}
\begin{itemize}
    \item As of June 24, 2021, 0.7\% (720K) of Egypt’s population is fully vaccinated at a speed of 86,831 doses per day \cite{Eg9}
    \item Population to Vaccine ratio: As of June 24, 2021, Egypt has vaccinated 4.04 people per 100 people. \cite{Ourworldindata}
\end{itemize}

\textbf{Vaccination Governance}: 
Egypt’s Ministry of Health is in charge of most vaccination protocols - the initial one has been to vaccinate 2 million Egyptians starting in February, with 4 million doses in agreement with Long Live Egypt (Tahya Misr) Fund and the Talaat Mostafa group. AstraZeneca is also receiving approval from the Egyptian Drug Authority. \cite{Eg6}

\subsubsection{Accessibility and Digital Solutions}
It is free of charge as a part of the Coronavirus Vaccine for Free initiative and for digital solutions - they are also launching an online registration portal (concurrently with Lebanon) that will allow citizens and healthcare workers to keep records of their vaccinations. Though Egypt’s portal does not make its methods and privacy protocols clear on its website, Lebanon’s ministry of health has assured that the platform protects privacy and has been tested for safety. The vaccine will be free of cost for all \cite{Eg4}. Egypt plans on acquiring a total of 4.5 million doses by June and has already vaccinated 149K individuals as of 4th April 2021. Egypt also plans on manufacturing the Sinopharm vaccine in its own facilities, and a negotiation deal with the pharmaceutical is in the works. Citizens can register through an online portal and opt for the vaccine. Egypt also received 800,000+ doses of the AstraZeneca vaccine from the COVAX facility \cite{Eg3}.

\subsubsection{Government Strategies and Decisions}
The emergency approval granted by Egypt was for frontline workers, and Egypt even had a publicity campaign describing the vaccine to be effective which a doctor considered to be propaganda. With two doses, a case study had also revealed that there was a 19 percent occurrence of dangerous side effects in recipients after the trial. Like Nepal, they do receive support from GAVI in acquiring the vaccine doses- however, for the frontline workers, Sinopharm has primarily aided in providing the vaccines to Egypt in the initial phases.

\subsubsection{Money spent}
\begin{itemize}
    \item Egypt received \$50 million from the World Bank Group's new Fast Track COVID-19 Facility in May 2020 as an emergency response.Egypt has been approved for \$50 million in financing, which is the highest amount available for a country with Egypt's population
\end{itemize}
\cite{Eg7}

\subsubsection{Variant info}
\begin{itemize}
    \item While the Indian strain of the virus has spread to 17 countries, including Iraq and Jordan, it has yet to be found in Egypt, according to the Ministry of Health for Preventive Affairs
\end{itemize}
\cite{Eg8}

\subsection{South Africa}

\subsubsection{Prioritization}

South Africa plans to vaccinate 40 million South Africans as a part of their vaccination roll out campaign. In this process of vaccination, the South African government has planned nationwide coverage  in three distinct phases. Included in this phased approach are various levels of prioritization groups of population. \cite{SA1}

\begin{itemize}
    \item  The first phase of vaccination will target the front-line health workers. The government is aware that there will be constrained vaccine supplies to cater to the demands during the initial phase of the implementation. So, it is planned to first provide vaccination to this priority group. As a part of the phase 1 plan, the government plans to vaccinate roughly 1.2 million front-line workers in the nation.

    \item In Phase 2, as the supply increases the following groups with a rough population of 16.6 million individuals will be targeted:
    \item As the supply of vaccines gradually increase, the country will progress to the phase 2 vaccination and plans to target the next proprietary groups of individuals. In this phase, they plan to vaccinate an approximate population of 16.6 million individuals.
     \begin{itemize} 
        \item This phase will include essential workers, persons in crowded settings, persons greater than 60 years of age (about 5 million people) and persons greater than 18 years of age who are suffering from associated comorbidities (about 8 million people) like diabetes, heart and kidney disease, HIV, Tuberculosis and other diseases that can make an individual immunocompromised.
        \item The essential workers who will be vaccinated as a part of the phase 2 will be officers in the police department, workers in the mining and security industry, staffs of retail food and grocery chains and shops and border control staff. In addition, they will also embrace staff of other public services including funeral places, academic and financial institutions, municipality and home affairs. The government targets to vaccinate 2.5 million people in this category.
        \item Individuals working in care homes, centers of detention, hospitality and tourism industry will also be covered in phase 2. As a part of this, it is targeted to vaccinate 1.1 million people. \cite{SA2}
     \end{itemize}
    \item Once the critical groups of the population are vaccinated, the final phase 3 will be implemented to cover approximately a population of 22.5 millions. In this final phase all other persons greater than 18 years of age will be vaccinated. This phase will be a continuous vaccination process, and the government has plans to shift into a part of the routine strategy.  \cite{SA2}
\end{itemize}

\subsubsection{Approved Vaccines}
The National Department of Health and the Ministerial Advisory committee discussed the various numbers of vaccines that will be needed to vaccinate the country’s populace. \cite{SA3}
These include:

\begin{itemize}
    \item Moderna
    \item Oxford/AstraZeneca
    \item Pfizer/BioNTech
    \item Johnson and Johnson
    \item Cipla
\end{itemize}

\subsubsection{Vaccine Administration and Distribution}
The government of South Africa wants to make sure that there is a supply of sufficient doses of vaccines to vaccinate its populace. This process collaborates with different vaccine suppliers and is working on supply chain logistics to ensure having enough quantities of vaccines in the shortest possible time. This will allow them to protect the most liable population in the community as early as possible. 
The government has also joined the COVAX programme, which is part of the Covid-19 Vaccines Global Access Facility's global pooled buying strategy. South Africa, in particular, has been promised that it will be among the first countries to receive a significant allotment of vaccines from this worldwide endeavour. \cite{SA4}

On February 1, 2021, the Serum Institute of India (SII) delivered 1 million doses of the Oxford University-AstraZeneca vaccine to South Africa. According to the National Institute for Communicable Diseases' website. \cite{SA3}
In the process of continuously procuring further supply the following has been done: 

\begin{itemize}
    \item The global CVOX programme is expected to produce around 12 million doses in total by March 2021.
    \item In conjunction with Johnson \& Johnson, it has also supplied 9 million immunisation doses. This will commence in the second quarter of 2021. Johnson \& Johnson has engaged the support of another company, Aspen, to assist in the production of these vaccines in the United States.

    \item  Pfizer is planned to supply another 20 million vaccine doses, with deliveries starting in the second quarter of 2021. \cite{SA3}
\end{itemize}
The country discovered a novel SARS-CoV-2 variation known as 501Y.V2 or B.1.351, which is now the virus's dominant form in the country. Following this, the government had to revisit the vaccine supply chain. The vaccine from AstraZeneca and the University of Oxford which the South African officials had begun to stockpile has proven largely ineffective at preventing mild-to-moderate COVID-19 due to B.1.351. In addition, it is mostly unknown if the vaccine offers any protection against severe disease caused by the new variants. \cite{madhi2021safety}

In Phase 1 of its countrywide vaccination drive, South Africa will employ the Johnson \& Johnson vaccine rather than the AstraZeneca vaccine. The 501Y.V2 has been confirmed to be resistant to the Johnson \& Johnson vaccination. As part of the Phase 3 Ensemble clinical trials, Johnson \& Johnson conducted efficiency experiments. The Sisonke Programme comes to mind in this context. This initiative is a research project to assess the Johnson \& Johnson vaccine's real-world efficacy among health care personnel. After vaccination, the programme will track, monitor, and assess the rate of hospitalizations and infection incidence among healthcare professionals. 
The South African Medical Research Council and the National Department of Health are collaborating on this programme. \cite{SA5}

\textbf{Vaccination Rate}
\begin{itemize}
    \item As of June 24, 2021, 0.8\% (479K) of Egypt’s population is fully vaccinated at a speed of 41,055 doses per day \cite{SA10}
    \item Population to Vaccine ratio: 4.3 vaccine doses administered per 100 people in South Africa as of June 24, 2021. \cite{Ourworldindata} 
\end{itemize}

\textbf{Vaccination Governance}: 
The Ministerial Advisory Committee (MAC) on Vaccines is the non-statutory advisory committee appointed by the Ministry of Health.  Another institutional agency in the country related to the vaccines safety and efficacy is South African Health Products Regulatory Authority (SAHPRA). This agency assesses the safety, efficacy and quality of vaccines used in the country using stringent assessment methods and mechanisms. \cite{SA6} \cite{SA3}

\subsubsection{Accessibility and Digital Solutions}
In the current context of the pandemic, the Government follows a centralised procurement policy in the sense that the government of South Africa will be the sole purchaser of vaccines. A national and provincial distribution network will then support this. A national rollout committee will monitor the vaccine  supply planning and implementation in both the public and private sectors.

It is planned to include all the vaccinated population into a national register and provide them with vaccination cards for tracking and monitoring the vaccinated population. 

The country has established an electronic vaccination data system (EVDS) to facilitate the nationwide vaccine rollout. The online EVDS is a self-enrolment platform where citizens can register for an appointment using a digital device. Citizens who meet the criteria will receive an SMS message with a unique code notifying them of the time and location. They will be required to present their national ID and a contact number to receive the immunisation. All Healthcare Workers (public \& private) who intend to be vaccinated in Phase I are encouraged to enroll on the EVDS. During the registration in the EVDS the following data is also collected and or provided:

\begin{itemize}
    \item Patient information (including demographic data, dose interval)
    \item Health establishment where service is accessible (name and type of clinic / hospital )
    \item Vaccine administered (manufacturer, batch number, etc.)
    \item Safety information of the vaccine administered (including Adverse Events Following Immunization)

    \item A record of vaccination issued to individuals where appropriate and required
\end{itemize}

\subsubsection{Government Strategies and Decisions}
Experts in the field have been appointed to a Ministerial Advisory Committee (MAC) on Covid-19 Vaccines. The committee is made up of vaccinology and virology experts, SAHPRA stakeholders, and WHO, Gavi Alliance, and Department of Health specialists.
They've devised a plan to ensure that immunizations are distributed fairly. The approach covers a variety of purchase techniques, financial implications, local production, opportunities, and determining immunisation priority groups.

\subsubsection{Money spent}
\begin{itemize}
    \item The South African government estimates a total funding of 19.3 billion Rand. It will be funded through budget allocations, contingency reserve, and emergency allocations
    \item The breakdown of the allocation is as follows:
    
    \begin{itemize}
        \item R6.5 billion will be allocated to the Department of Health to procure and distribute vaccines
        \item R100 million will be allocated to the South African Medical Research Council for vaccine research
        \item R2.4 billion will be allocated to provincial health departments to administer vaccines
        \item R50 million will be allocated to the Government Communication and Information System for a communications campaign on vaccines
    \end{itemize}
    
    \item Vaccines will be provided free of charge according to rollout schedules \cite{SA7}
    \item Over the next two years, South Africa will allocate more than R10 billion (\$712 million) to purchase and deliver vaccines \cite{SA8}
\end{itemize}

\subsubsection{Variant info}
\begin{itemize}
    \item South Africa is one of the African continent's hardest impacted countries. The B.1.617.2 and B.1.1.7 variations have been discovered, according to the National Institute of Communicable Diseases
    \item The first variant, B.1.617.2 (initially detected in India), has been detected in four positive cases, the institute said. Two of those are in Gauteng, South Africa’s most populous province
    \item The second variant, B.1.1.7 (initially detected in the U.K.), has been detected in 11 cases
\end{itemize}
\cite{SA9}

\subsection{Estonia}

\subsubsection{Prioritization}
Estonia has started mass vaccinations, and commencing from May 17 all Estonian people can get vaccinated against COVID-19.  When sufficient vaccine quantities are available, the government plans to  open up the possibility of vaccinating everyone who wants to get vaccinated. The government is planning to provide vaccines for free to all individuals living in Estonia. This will also include those individuals who do not have an official health insurance. 

Because vaccine supplies may be limited, the state began immunising healthcare workers, residents and employees of social care facilities, and people who are considered high-risk because to their elderly age.

In order to ensure that society's vital services continue to run, Estonia adopts a phased vaccination approach, with initial priority given to people in a high-risk group or who need to be vaccinated first.
In collaboration with the Estonian Family Doctors' Association, the country's Immunoprophylaxis Expert Committee has created six risk groups and divided diseases into two categories: very high risk and high risk. Prioritization will follow the following order:

\begin{itemize}
    \item Group 1 includes all adults over the age of 80
    \item Group 2 People who are 70 years old or older and have concomitant diseases or disorders that put them at a very high or high risk of serious disease progression.

    \item People who are 70 years old or older, regardless of comorbidities, fall into group 3. This group also includes people aged 16 to 69 who have diseases or conditions that put them at a high risk of serious disease progression.

    \item Individuals in group 4 are those who are 60 years old or older and have diseases or disorders that put them at a high risk of serious disease progression.

    \item People in Group 5 are those who are 50-59 years old or older and have diseases or disorders that put them at a high risk of serious disease progression.

    \item People in the last group are 16-49 years old or older and have diseases or disorders that put them at a high risk of serious disease progression.
\end{itemize}

\cite{Estonia1}
\cite{Estonia2}

\subsubsection{Approved Vaccines}
 The following vaccines have been approved for use in the country:

\begin{itemize}
    \item Moderna
    \item Vaxzevria (COVID-19 Vaccine AstraZeneca)
    \item Pfizer/BioNTech
    \item Johnson and Johnson
\end{itemize}

\subsubsection{Vaccine Administration and Distribution}
The process of vaccine administration started with a pre-analysis of all the patient lists of all the family doctors in the country. The Estonian Health Insurance Board mapped people who fall into a COVID-19 risk group based on diseases and illnesses defined by the state Immunoprophylaxis Expert Committee as part of this process. Patients were categorised by the Health Insurance Board based on diagnosis codes found on medical bills and prescriptions written in the prior five years. Following the preparation of the lists, the family doctors are given the opportunity to review them and make any necessary modifications. Starting March 22, members of risk groups will be able to get information about their risk group status through the patient portal.

Estonia is participating within the joint EU procurement for purchasing the coronavirus vaccines. The joint European Union vaccine portfolio contains vaccines and vaccine candidates of seven different vaccine manufacturers. The European Commission has signed pre-purchase agreements with the following vaccine manufacturers – AstraZeneca, Sanofi, Janssen Pharmaceutica NV, Pfizer/BioNTech, Curevac and Moderna. Estonia has also joined pre-purchase agreements with five vaccine manufacturers - AstraZeneca, Janssen Pharmaceutica NV, Pfizer/BioNTech, Curevac and Moderna. Additionally, negotiations are underway between the European Commission and the vaccine manufacturer Novavax and  Valneva.
\cite{Estonia3}

\textbf{Vaccination Rate}
\begin{itemize}
    \item As of June 24, 2021, 27.57\% (365K) of Estonia’s population is fully vaccinated at a speed of 8,757 doses per day \cite{Estonia11}
    \item Population to vaccine ratio: 68.95 doses administered per 100 people as of June 24, 2021 \cite{Ourworldindata}
\end{itemize}

\textbf{Vaccination Governance}: 
The Estonian Health Board and the Ministry of Social Affairs govern the country's overall process. Partial redistribution of vaccines and delivery schedules are managed by the Steering Committee for the Joint Procurement of COVID-19 Vaccines.

\cite{Estonia4}

\subsubsection{Accessibility and Digital Solutions}
The central patient portal of the national digital registry allows the citizens to see his or her health data, and to book specialist appointments. In addition to the national digital registry, the country also has a central national database for their Health Information System and this allows the healthcare professionals to notify the older and younger people about their risk grouping. The country is collaborating with WHO in implementing VaccineGuard a digital platform to effectively support vaccination. VaccineGuard is a digital platform that connects various healthcare stakeholders and the different participants in the vaccine ecosystem of the country. Vaccine producers, hospitals, public health authorities, certificate suppliers, wellness app providers, citizens, border guards, vaccination programme managers, and insurance companies are among the numerous actors in this ecosystem. VaccineGuard will promote and enable these participants to share and verify data across organisational and international boundaries. VaccineGuard will also enable the development of real-time insights, pharmacovigilance, counterfeit identification, and a slew of other features that will help the government respond to pandemics faster and more reliably.

\cite{Estonia5}
\cite{Estonia6}
\cite{Estonia7}

\subsubsection{Government Strategies and Decisions}
The Estonian government has strategically signed a memorandum of understanding with WHO that will allow Estonian entrepreneurs and researchers to start working on digital immunization certificate, interoperability, and other projects. This will allow for an improved immunization record, robust and transparent exchange of vaccination data across borders of what we hope will be an eventual vaccine for COVID-19, as well as immunizations records for other diseases.
\cite{Estonia8}

\subsubsection{Money spent}
\begin{itemize}
    \item Estonia is planning to donate 800,000 vaccines through the COVAX initiative and to the EU’s Eastern Partnership countries (Georgia, Ukraine, Moldova, Armenia, Azerbaijan, and Belarus). Additionally, Estonia will donate €70,000 to COVAX
\end{itemize}
\cite{Estonia9}

\subsubsection{Variant info}
\begin{itemize}
    \item As of May 6, 2021, 90\% of COVID-19 cases are the British variant (B.1.1.7)
\end{itemize}
\cite{Estonia10}

\subsection{Kazakhstan}

\subsubsection{Prioritization}
On February 1, 2021, the vaccination with the Russian Sputnik V vaccine began with a supply of 20,000 units.

The vaccination will be carried out in stages, taking into account the supply of the vaccine.
The vulnerable groups of the population at a high risk of infection and spread of coronavirus will be vaccinated first. Next, the rest of the population will be vaccinated.

In detail, the following groups of the population will be vaccinated at the first stage:

\begin{itemize}
    \item healthcare workers of infectious diseases hospitals
    \item ambulance
    \item intensive care units
    \item emergency hospitals
    \item hospital admissions
    \item sanitary and epidemiological service
\end{itemize}

The second stage of vaccination will include:

\begin{itemize}
    \item teachers of secondary schools and universities
    \item other healthcare workers
\end{itemize}

The third stage of vaccination will include:

\begin{itemize}
    \item teachers of boarding schools and preschool institutions
    \item students
    \item persons with chronic diseases
\end{itemize}

Additional vulnerable sections of the population, such as employees of the Ministry of Emergency Situations, Ministry of Internal Affairs, Ministry of Defense, National Security Committee, State Security Service, and others, will be vaccinated in the future.
If the vaccinations are registered and available on the market, it is predicted that around six million individuals will be immunised by the end of the year.

\cite{Kaza1}
\cite{Kaza2}

\subsubsection{Approved Vaccines}
 Approved Vaccines in Kazakhstan

\begin{itemize}
    \item Sputnik V vaccine
    \item QazVac vaccine - No clear evidence of the start of usage of this locally manufactured vaccine
\end{itemize}

\cite{Kaza3}

\subsubsection{Vaccine Administration and Distribution}
Kazakhstan citizens will receive free vaccination within the guaranteed volume of free healthcare. According to the Ministry of Health, about \$118.5 million will be allocated from the budget to vaccinate Kazakhstan citizens against coronavirus.

\cite{Kaza4}
\cite{Kaza5}

\textbf{Vaccination Rate}
\begin{itemize}
    \item A cumulative of 4.71 million doses have been administered in the country
    \item 9.55\% of the population has been fully vaccinated, while 15.55\% has gotten at least one dose of immunisation. At least one dose of vaccine has been administered to a total of 2.92 million persons. \cite{Kaza6}
    \item According to the platform OurWorldinData, the number of administered coronavirus vaccine doses per 100 people in Kazakhstan is 25.10 (as of June 24, 2021). This is considered as a single dosage, and depending on the dose regimen, it may or may not equal the total number of people vaccinated (e.g. people receive multiple doses). This is high compared to neighboring countries like Uzbekistan(8.05 per 100 people) and Kyrgyzstan (2.66 per 100 people)
\end{itemize}
\cite{Ourworldindata}

\textbf{Vaccination Governance}: 
Vaccination of the population is carried out at the vaccination points of territorial medical organizations. Vaccination of a special contingent is carried out at the duty point of the relevant department. In some cases (small population, lack of stationary inoculation room, lack of refrigeration equipment with low temperature), mobile vaccination points or mobile vaccination groups are organized for vaccination.

Vaccination group is formed taking into account the daily load per team at each vaccination point - no more than 60 vaccinations (on-site or mobile - no more than 40 vaccinations). 

In this case, 1 vaccination team consists of: 1 doctor, 1 vaccinating nurse and, if necessary, 1 registrar

In rural areas, it is allowed to create a vaccination team consisting of 1 paramedic (in the absence of a doctor), 1 vaccination nurse and, if necessary, 1 registrar.

Vaccination is carried out in the prescribed registration forms with indication of the date of vaccination, type of vaccination (component I or II), manufacturer, serial number, reaction to the vaccine (personal logbook of prophylactic vaccinations against KVI, preventive vaccination card, outpatient medical record, vaccination department of the medical information system).

\subsubsection{Accessibility and Digital Solutions}
Electronic vaccination passport is a document certifying the receipt of preventive vaccinations.  It contains information about what medications a person has taken, what dose, what days, and what symptoms have occurred after vaccination. After a person receives the coronavirus vaccine, information about him will be available in the eGov mobile application.

\subsubsection{Government Strategies and Decisions}
In Kazakhstan, the Russian vaccine is produced at the Karaganda Pharmaceutical Complex. More precisely, the substance is imported from Russia, and in Kazakhstan the finished substance is made from that substance. Kazakhstani materials and raw materials are used

\subsubsection{Money spent}
\begin{itemize}
    \item TBD
\end{itemize}

\subsubsection{Variant info}
\begin{itemize}
    \item TBD
\end{itemize}

\subsection{Nepal}

\subsubsection{Prioritization}
Nepal’s Inoculation strategy is contingent on its ability to acquire more doses from other countries. Out of the various vaccines currently approved, Nepal began its vaccination program in the first phase by vaccinating with the Astrazeneca COVISHIELD®.

\begin{itemize}
    \item (Highest Priority) Nepal began its vaccination program in the first phase by vaccinating front-line medical workers, sanitation workers, and security officers
    \item (High Priority) After the completion of the initial phase, and as part of the second phase, journalists, diplomats, and government employees have been prioritized
    \item (High Priority) Nepal has also set age as a criterion for prioritization. Starting from March 7th, the vaccination drive has primarily focused on persons over 55 years of age
    \item Everyone else will be vaccinated later on, after the completion of the above categories \cite{Nepal2}
\end{itemize}

\subsubsection{Approved Vaccines}
 Approved Vaccines in Nepal \cite{Nepal1}

\begin{itemize}
    \item Oxford/AstraZeneca
    \item China/Sinopharm
    \item India/Bharat BioTech Covaxin
\end{itemize}

\subsubsection{Vaccine Administration and Distribution}
In the first phase of the program, district hospitals undertook the responsibility of distribution. Owing to a less than ideal rate of vaccination, Nepal then set up primary health centers to aid in the process. Nepal has set up 177 medical centers to carry out the program. \cite{Nepal10} In the capital city of Kathmandu itself, over 15 medical facilities, comprising over 32 medical wards have been set up since the second vaccination phase. Eligible persons have been asked to carry their personal identification card, proof of their citizenship, and conclusive proof of their occupation. Education staff like professors and teachers are required to bring their identity cards, whereas vehicular operators are required to bring a letter ascertaining their position as the same. \cite{Nepal3}

\textbf{Vaccination Rate}
\begin{itemize}
    \item As of June 24, 2021, 2.51\% (731K) of Nepal’s population is fully vaccinated at a speed of 22,003 doses per day \cite{Nepal15}
    \item 11.09 doses administered per 100 people as of June 24, 2021 \cite{Ourworldindata}
\end{itemize}

\textbf{Vaccination Governance}: 
The Ministry of Health and Population (MoHP) carried out the immunization program in Nepal. Since Nepal has largely depended on vaccine supplies from China, India and the GAVI framework, its government has had to work in close proximity with international leaders and representatives to acquire vaccine jabs \cite{Nepal8}. Nepal has negotiated a deal of 5 million doses of the COVISHIELD® vaccine with the Serum Institute of India which is yet to be finalised. Amidst India’s reduction in vaccine exports to cater to its domestic population in the second wave, many countries including Nepal have suffered. Foreign Minister of Nepal, Mr Pradeep Gyawali has remarked that Nepal has already paid for the second round of 1 million vaccine doses from India, and has been forced to halt its immunization program. \cite{Nepal9}

\subsubsection{Accessibility and Digital Solutions}
As of 9th April 2021, Nepal ran out of the COVISHIELD® vaccine doses. Though the vaccine is free of cost, its distribution has remained limited. Nepal has not been able to meet its immunization goals so far, but with a new resurgence of daily cases, it is said to have started its immunization again with China’s vaccines. \cite{Nepal8} Nepal’s use of technology in fighting the COVID-19 pandemic, and its immunization program has been limited at best. As of 9th April 2021, there is no digital system for vaccination-appointment, and all scheduling and organising is done on-site. Furthermore, Nepal has loosely been tracking the number of doses that it has administered, and has not updated its COVID-19 vaccination data in the last 2 weeks. \cite{Nepal7}

\subsubsection{Government Strategies and Decisions}
Nepal also became the first country in the Asia-Pacific region to administer vaccines to refugees. As of 24th March, over 700 refugees who are also senior persons have been administered the first dose.\cite{Nepal4} Nepal financing for the inoculation drive has largely depended on charitable funds from the World-Bank, and vaccination supplies free of cost from China and India. Nepal has been the beneficiary of a \$75 million donation from the World Bank, over 800,000 shots of the Sinopharm vaccine from China \cite{Nepal5}, and over 450,000 jabs from India.\cite{Nepal6}

\subsubsection{Money spent}
\begin{itemize}
    \item Nepal received \$75 million from the World Bank for better “access to safe and effective COVID-19 vaccines and equitable vaccine distribution”. This additional funding “builds on the \$29 million for the original COVID-19 Emergency Response and Health Systems Preparedness Project \cite{Nepal11} that was approved in April 2020.”
    \item Here is a rough breakdown of how the fund will be spent \cite{Nepal12}:
    
        \begin{itemize}
            \item Most (90\%) will be used for purchasing COVID‐19 vaccines and deploying them to prioritized populations
            \item Rest will be used to support the COVID‐19 vaccination effort and procure other COVID‐19-related supplies such as diagnostic tests, laboratory equipment and therapeutics
        \end{itemize}
        
    \item The Nepalese government has also taken a \$250 million concessional loan from the Asian Development Bank (ADB). This is in addition to a \$300,000 grant received from ADB to procure medical supplies \cite{Nepal13}
\end{itemize}

\subsubsection{Variant info}
\begin{itemize}
    \item According to Nepal's Ministry of Health and Population, three variants of the virus have been detected.  In 97 percent of the samples, the B.1.617.2 variant was found, while the B.1.617.1 variant was found in the rest
\end{itemize}
\cite{Nepal14}

\subsection{Republic of Korea}

\subsubsection{Prioritization}
The priority groups are as follows:

\begin{itemize}
    \item Healthcare workers
    \item First responders, contact tracers
    \item Workers at care facilities for the elderly
    \item Senior residents

\end{itemize}

\subsubsection{Approved Vaccines}
 The Republic of Korea has signed deals to purchase COVID-19 vaccines from the following Approved Vaccines:  

\begin{itemize}
    \item Pfizer-BioNTech
    \item Moderna (NIAID)
    \item AstraZeneca-Oxford (AZD1222)
    \item Johnson \& Johnson’s (JNJ-78436735)
    \item Sputnik V.
\end{itemize}

\subsubsection{Vaccine Administration and Distribution}
The government of the Republic of Korea started its first wave of vaccinations in February of 2021. Korea has demarcated approximately 250 vaccine centers alongside the tens of thousands of hospitals as facilities that carry out the program. The first group to receive the vaccine were approximately 50,000 frontline COVID-19 healthcare workers throughout February. The next group to receive the vaccine were the first responders, residents, contact tracers, and workers at care facilities for the elderly. Korea plans to immunize the third group, the elderly people over 65, homeless people, and the remainder of the previous group during May \& June. From July to November, the government aims to vaccinate the general public with bulk use of the Pfizer-BioNTech and Moderna vaccines to achieve herd immunity by the end of the year. These groups serve as a rough estimate for the government’s ambitious goal of inoculating 70\% of its ~52,000,000 people.

\textbf{Vaccination Rate}
\begin{itemize}
    \item As of June 24, 2021, 8.82\% (4.52M) of Korea’s population is fully vaccinated at a speed of 332,606 doses per day \cite{SKorea9}
    \item South Korea has vaccinated 36.27 people for every 100 people, as of June 24, 2021. \cite{Ourworldindata}
\end{itemize}

\textbf{Vaccination Governance}: 
The government’s ministry of health and welfare is responsible for proper functioning of the program. The ministry of Drug Safety is responsible for testing the safety and efficacy of all the vaccines.The government is implementing a two-track strategy that involves using the COVAX Facility to distribute 10 million doses of vaccination in addition to acquisition of 20 million more from AstraZeneca. The strategy is to have the first and second dosage of COVID-19 vaccine 2 months apart in order to observe potentially negative side effects. The government has also secured 106 million doses as of January 4, 2021, allowing for the full vaccination of 56 million people.  It recently decided to exclude people under 30 from the Astrazeneca COVISHIELD drive due to blood clotting concerns.  A leading pharmaceutical in South Korea, Huons Global, with the government's support plans on manufacturing over 100 million doses of the Sputnik V vaccine per month.

\subsubsection{Accessibility and Digital Solutions}
COVID-19 vaccines for all of South Korea’s 52 million people are free of cost \cite{SKorea2}. South Korea will issue a ‘green-pass’ to all of its vaccinated residents via a mobile-app to ensure that fully vaccinated people can return to normalcy. The vaccination digital passports are protected via a state of the art blockchain technology to prevent counterfeiting \cite{SKorea3}. The government’s responses to disease control, including vaccinations and a responsive map of the state of COVID-19 in South Korea, are made available through the Coronomap website maintained by the government. \cite{SKorea4}

\subsubsection{Government Strategies and Decisions}
South Korea won’t provide its residents the opportunity of choosing a vaccine of their own. The vaccines will be voluntary, however, the government aims to encourage as many people as possible to reach the herd-immunity target. As of now, the government denies any causal relationship between injuries and the COVID-19 vaccines. It is unclear as of now the extent of customer protection and compensation that will be offered if and when a relationship between the recent deaths and the vaccine is proved. \cite{SKorea5}

\subsubsection{Money spent}
\begin{itemize}
    \item The exact amount of money the South Korean government has spent on vaccine procurement remains undisclosed
    \item “South Korea's vaccine rollout has been hampered by global shortages and shipment delays, while the government has also sought to boost immunisation by easing restrictions for people who have been vaccinated.” \cite{SKorea6}
    \item US donated 1 million doses of Johnson \& Johnson’s (J\&J) vaccines to Korea, including 550,000 shots for the South Korean troops \cite{SKorea7}
\end{itemize}

\subsubsection{Variant info}
\begin{itemize}
    \item All four variant strains (those initially detected in the UK, India, South Africa, and Brazil) are present in Korea. Variant cases are increasing at an alarming rate
    \item As of May 12, 2021, the rate for estimating the spread of major COVID-19 variants (UK, South African, Brazilian) within South Korea has risen to nearly 25\%, nearly double that of last week's 12.8\%
\end{itemize}
\cite{SKorea8}

\subsection{France}

\subsubsection{Prioritization}
Priority groups for vaccine administration have majorly been divided on the basis of two factors: age and pre existing conditions that increase the vulnerability of patients getting seriously affected by COVID-19. At the beginning phase of vaccine roll out in January, people above the age of 75 or people having pre existing conditions which could lead to serious effects (irrespective of age) were allowed to get vaccinated. Since March the age limit of people allowed to get vaccinated changed to 70 years old which further changed to 55 years old in April 2021\cite{F1}. Currently the Priority groups allowed to get vaccinated are as follows:

\begin{itemize}
    \item People aged 55 and over
    \item People who are residents in residential facilities for elders or hosted in residential autonomy
    \item People over 18 years of age with pre existing conditions that can lead to severe form of COVID-19 such as:
    \begin{itemize}
        \item suffering from multiple chronic pathologies and presenting at least two organ insufficiencies
        \item transplanted by allogeneic hematopoietic stem cell transplantation
        \item Down’s Syndrome
        \item with severe chronic kidney disease, including dialysis patients
        \item Cancer
        \item Malignant haematological diseases during treatment with chemotherapy
        \item People having disabilities in specialised homes or foster homes medicalised
    \end{itemize}
    \item Residents 60 and over in homes of migrant workers
    \item Health sector professionals and people working in the healthcare industry like:
    \begin{itemize}
        \item medical secretaries in town offices and medical assistants
        \item professions with "use of title" recognized by various uncoded laws (osteopaths, chiropractors, psychotherapists, psychologists)
        \item home help professionals and employees of the private employer working with vulnerable elderly and disabled people (receiving APA or PCH)
        \item the personnel making up the crews of the vehicles of medical transport companies
        \item medical regulation assistants during their internships in an establishment or in SMUR
        \item veterinarians
        \item professionals in serviced residences
        \item professionals in specialized accommodation centers for people with Covid-19
        \item professional and volunteer firefighters
        \item service providers and distributors of equipment working in patients' homes
        \item Medical biologists
        \item Pharmacist
        \item medical physicist
        \item  Medical auxiliaries, nursing assistants, childcare auxiliaries and ambulance attendants
        \item general or specialized nurse, advanced practice nurse
        \item Physiotherapist
        \item Chiropodist
        \item occupational therapist and psychomotor therapist
        \item Paramedic
        \item dental assistant.
        \item prosthetist and orthotist for equipment for disabled people
        \item medical electro radiology manipulator
        \item anti-Covid mediators
        \item professions with "use of title" recognized by various uncoded laws (osteopaths, chiropractors, psychotherapists, psychologists)\cite{F2}
    \end{itemize}
\end{itemize}

\subsubsection{Approved Vaccines}
 Four vaccines have been approved for administration in France.\cite{F3}

\begin{itemize}
    \item Pfizer-BioNTech
    \item Moderna (NIAID)
    \item AstraZeneca-Oxford (AZD1222)
    \item Johnson \& Johnson’s (JNJ-78436735)
\end{itemize}

\subsubsection{Vaccine Administration and Distribution}
18.3 million doses have been administered till now and the result is 5.03 million people being fully vaccinated in France\cite{Ourworldindata}. Two distinct logistics circuits were set up to ensure that France’s nursing homes (around 10,000) receive the vaccines with ease, including the Pfizer-BioNTech jabs, which require conditions like storage temperatures of -70 degrees Celsius. The vaccines would be stored in assigned pharmaceutical warehouses before reaching nursing homes and hospitals.  \cite{F4}

\textbf{Vaccination Rate}
\begin{itemize}
    \item 51 million doses given
    \item 18.57 million fully vaccinated
    \item 27.49\% of the population fully vaccinated
    \item Population to vaccine ratio: 75.47 doses administered for every 100 people as of June 24, 2021. \cite{Ourworldindata}
\end{itemize}

\textbf{Vaccination Governance}: 
The government has appointed Professor Alain Fischer, a paediatric immunologist, to lead France's vaccine plan. The French National Authority of Health has made vaccination rollout and distribution plans recommendations. \cite{F5}

\subsubsection{Accessibility and Digital Solutions}
France is participating in a month-long trial of an air travel vaccine passport. On Thursday, transport minister Jean-Baptiste Djebbari announced the start of a month-long pilot for Air France passengers (11 March). Passengers traveling to Martinique and Guadeloupe in the French Caribbean will be able to use the International Chamber of Commerce's (ICC) AOKpass app citeF8 to verify a COVID-19 immunisation certificate negative test result under the scheme. \cite{F9}
France became the first EU country to test an app-based travel permit that retains negative COVID-19 test results, and vaccination certificates will soon be accepted on flights to Corsica and its offshore territories.  \cite{F10}

\subsubsection{Government Strategies and Decisions}
The French government has stepped up immunisation efforts and developed policies and strategies to protect specific priority groups based on pre-existing conditions and age. The government has implemented laws that prioritise health-care workers, the elderly, and persons with medical disorders that put them at risk of acquiring a severe form of COVID-19 as those who must be vaccinated first. France has announced a multi-stage vaccination policy, with priority given to those who are at risk. The majority of patients who died in France due to COVID-19 (approximately 78 percent) were over 75 years old. They currently occupy 28 percent of ICU beds, making senior citizens the first category at danger from the virus. \cite{F6} French government has consulted private firms like Mckinsey \& Company for vaccine distribution strategies, a move which has been heavily criticised.\cite{F7}
French government has consulted various health professionals and vaccine manufacturers for the plan of action.

\subsubsection{Money spent}
\begin{itemize}
    \item France has set aside 1.5 billion euros (\$1.82 billion) from this year's social security budget to cover the expense of COVID-19 vaccines for all members of its social security system.
\end{itemize}
\cite{F11}

\subsubsection{Variant info}
\begin{itemize}
    \item According to French Health Minister Olivier Veran, some 20 persons in France have been diagnosed with the B.1.617 strain
\end{itemize}
\cite{F12}

\section{Conclusion and Future Scope}
The world needs safe, effective COVID-19 vaccine strategies to return to a pre-covid normal situation. The vaccination process is getting adopted in every country across the globe, and there are certain challenges in terms of logistics, distribution, availability, and accessibility of vaccines across various countries. This study aims to help different countries learn from each other to develop and improve their vaccine roll-out plans based on different parameters. This study indicates the various strategies, distribution, and challenges faced in different countries for COVID-19 vaccinations. It also shared the information related to the spread of variants in different parts of the world and how the vaccination rate is curbing the rate of infection spread. 
We also focused on governance strategies and accessibility of the vaccination information and supply of the same. 
Vaccinations have been proven helpful to provide herd immunity in certain countries and have not been effective in countries with large populations and a shortage of vaccines. 
The future version of this study will focus on post-vaccination's social and economic effects and returning to normalcy.

\subsection*{Acknowledgments}
We are grateful to Riyanka Roy Choudhury, CodeX Fellow, Stanford University, Adam Berrey, CEO of PathCheck Foundation, Dr. Brooke Struck, Research Director at The Decision Lab, Canada, Vinay Gidwaney, Entrepreneur and Advisor, PathCheck Foundation, and Paola Heudebert, co-founder of Blockchain for Human Rights, Alison Tinker, Saswati Soumya, Sunny Manduva, Bhavya Pandey, and Aarathi Prasad for their assistance in discussions, support, and guidance in the writing of this paper.
\bibliography{iclr2021_conference}

\begin{thebibliography}{209}
\providecommand{\natexlab}[1]{#1}
\providecommand{\url}[1]{\texttt{#1}}
\expandafter\ifx\csname urlstyle\endcsname\relax
  \providecommand{\doi}[1]{doi: #1}\else
  \providecommand{\doi}{doi: \begingroup \urlstyle{rm}\Url}\fi

\bibitem[Acuna-Zegarra et~al.(2020)Acuna-Zegarra, Diaz-Infante, Baca-Carrasco,
  and Liceaga]{acuna2020covid}
Manuel~Adrian Acuna-Zegarra, Saul Diaz-Infante, David Baca-Carrasco, and
  Daniel~Olmos Liceaga.
\newblock Covid-19 optimal vaccination policies: a modeling study on efficacy,
  natural and vaccine-induced immunity responses.
\newblock \emph{medRxiv}, 2020.

\bibitem[Agency(2020{\natexlab{a}})]{Russ9}
Anadolu Agency.
\newblock Russia to hold phase 3 of covid-19 vaccine trial abroad,
  2020{\natexlab{a}}.
\newblock URL
  \url{https://www.aa.com.tr/en/latest-on-coronavirus-outbreak/russia-to-hold-phase-3-of-covid-19-vaccine-trial-abroad/1912694}.

\bibitem[Agency(2021{\natexlab{a}})]{Kaza3}
Anadolu Agency.
\newblock Kazakhstan's covid-19 vaccine to be bottled in turkey,
  2021{\natexlab{a}}.
\newblock URL
  \url{https://www.aa.com.tr/en/asia-pacific/kazakhstans-covid-19-vaccine-to-be-bottled-in-turkey/2204040}.

\bibitem[Agency(2021{\natexlab{b}})]{Turkey16}
Anadolu Agency.
\newblock Turkey: 85\% of new covid-19 cases due to uk variant,
  2021{\natexlab{b}}.
\newblock URL
  \url{https://www.aa.com.tr/en/health/turkey-85-of-new-covid-19-cases-due-to-uk-variant/2206883}.

\bibitem[Agency(2020{\natexlab{b}})]{It8}
Italian~Medicines Agency.
\newblock First aifa report on covid-19 vaccine surveillance,
  2020{\natexlab{b}}.
\newblock URL
  \url{https://www.aifa.gov.it/en/web/guest/-/primo-rapporto-aifa-sulla-sorveglianza-dei-vaccini-covid-19}.

\bibitem[Agency(2020{\natexlab{c}})]{Russ1}
Russian~News Agency.
\newblock Russian health ministry expands list of priority coronavirus
  vaccination groups, 2020{\natexlab{c}}.
\newblock URL \url{https://tass.com/coronavirus-outbreak-in-china/1197127}.

\bibitem[Asia(2021{\natexlab{a}})]{SKorea7}
Channel~News Asia.
\newblock South korea says 1 million doses of j\&j covid-19 vaccines to arrive
  this week from us, 2021{\natexlab{a}}.
\newblock URL
  \url{https://www.channelnewsasia.com/news/asia/south-korea-says-1-million-doses-of-j-j-covid-19-vaccines-to-arrive-this-week-from-us-14914688}.

\bibitem[Asia(2021{\natexlab{b}})]{Singapore5}
Channel~News Asia.
\newblock Covid-19 vaccination begins for land transport sector with jabs to be
  offered to 80,000 workers, 2021{\natexlab{b}}.
\newblock URL
  \url{https://www.channelnewsasia.com/news/singapore/covid-19-vaccination-drive-begins-for-land-transport-workers-lta-14038186}.

\bibitem[Asia(2021{\natexlab{c}})]{TH4}
Nikkei Asia.
\newblock Thai businesses to offer covid shots to wealthy and companies,
  2021{\natexlab{c}}.
\newblock URL
  \url{https://asia.nikkei.com/Spotlight/Coronavirus/COVID-vaccines/Thai-businesses-to-offer-COVID-shots-to-wealthy-and-companies}.

\bibitem[Augustin(2021)]{Malaysia2}
Robin Augustin.
\newblock What’s our covid-19 vaccination game plan, ask mps, 2021.
\newblock URL
  \url{https://www.freemalaysiatoday.com/category/nation/2021/01/31/whats-our-covid-19-vaccination-game-plan-ask-mps/}.

\bibitem[Authority(2021{\natexlab{a}})]{Singapore11}
Health~Sciences Authority.
\newblock Suspected vaccine adverse event (vae) online reporting form,
  2021{\natexlab{a}}.
\newblock URL
  \url{https://eservice.hsa.gov.sg/adr/adr/vaeOnline.do?action=load}.

\bibitem[Authority(2021{\natexlab{b}})]{SA6}
South African Health Products~Regulatory Authority.
\newblock Covid-19 vaccines regulatory status update, 2021{\natexlab{b}}.
\newblock URL
  \url{https://www.gov.za/sites/default/files/gcis_documents/SAHPRA_COVID-19%20vaccines%20update.pdf}.

\bibitem[Baden et~al.(2021)Baden, El~Sahly, Essink, Kotloff, Frey, Novak,
  Diemert, Spector, Rouphael, Creech, et~al.]{baden2021efficacy}
Lindsey~R Baden, Hana~M El~Sahly, Brandon Essink, Karen Kotloff, Sharon Frey,
  Rick Novak, David Diemert, Stephen~A Spector, Nadine Rouphael, C~Buddy
  Creech, et~al.
\newblock Efficacy and safety of the mrna-1273 sars-cov-2 vaccine.
\newblock \emph{New England Journal of Medicine}, 384\penalty0 (5):\penalty0
  403--416, 2021.

\bibitem[Bae et~al.(2020)Bae, Gandhi, Kothari, Shankar, Bae, Patwa, Sukumaran,
  TV, Yu, Misra, et~al.]{baechallenges}
Joseph Bae, Darshan Gandhi, Jil Kothari, Sheshank Shankar, Jonah Bae, Parth
  Patwa, Rohan Sukumaran, Sethuraman TV, Vanessa Yu, Krutika Misra, et~al.
\newblock Challenges in equitable covid-19 vaccine distribution: A roadmap for
  digital technology solutions.
\newblock \emph{arXiv preprint arXiv:2012.12263}, 2020.

\bibitem[Bae et~al.(2021{\natexlab{a}})Bae, Sukumaran, Shankar, Sharma, Singh,
  Nazir, Kang, Srivastava, Patwa, Singh, et~al.]{bae2021mobile}
Joseph Bae, Rohan Sukumaran, Sheshank Shankar, Anshuman Sharma, Ishaan Singh,
  Haris Nazir, Colin Kang, Saurish Srivastava, Parth Patwa, Abhishek Singh,
  et~al.
\newblock Mobile apps prioritizing privacy, efficiency and equity: A
  decentralized approach to covid-19 vaccination coordination.
\newblock \emph{arXiv preprint arXiv:2102.09372}, 2021{\natexlab{a}}.

\bibitem[Bae et~al.(2021{\natexlab{b}})Bae, Sukumaran, Shankar, Srivastava,
  Iyer, Mahindra, Mirza, Arseni, Sharma, Agrawal, et~al.]{bae2021safepaths}
Joseph Bae, Rohan Sukumaran, Sheshank Shankar, Saurish Srivastava, Rohan Iyer,
  Aryan Mahindra, Qamil Mirza, Maurizio Arseni, Anshuman Sharma, Saras Agrawal,
  et~al.
\newblock Mit safepaths card (misaca): Augmenting paper based vaccination cards
  with printed codes.
\newblock \emph{arXiv preprint arXiv:2101.07931}, 2021{\natexlab{b}}.

\bibitem[Bank(2020{\natexlab{a}})]{Nepal13}
The Asian~Development Bank.
\newblock Adb approves \$250 million support for nepal's covid-19 response,
  2020{\natexlab{a}}.
\newblock URL
  \url{https://www.adb.org/news/adb-approves-250-million-support-nepals-covid-19-response}.

\bibitem[Bank(2020{\natexlab{b}})]{Eg7}
The~World Bank.
\newblock Egypt: World bank provides us\$ 50 million in support of coronavirus
  emergency response under new fast-track facility, 2020{\natexlab{b}}.
\newblock URL
  \url{https://www.worldbank.org/en/news/press-release/2020/05/17/egypt-world-bank-provides-us-50-million-in-support-of-coronavirus-emergency-response-under-new-fast-track-facility}.

\bibitem[Bank(2021{\natexlab{a}})]{Nepal11}
The~World Bank.
\newblock Nepal: Covid-19 emergency response and health systems preparedness
  project, 2021{\natexlab{a}}.
\newblock URL
  \url{https://projects.worldbank.org/en/projects-operations/project-detail/P173760}.

\bibitem[Bank(2021{\natexlab{b}})]{Nepal12}
The~World Bank.
\newblock Nepal receives \$75 million for covid-19 vaccines and stronger
  response to pandemic, 2021{\natexlab{b}}.
\newblock URL
  \url{https://www.worldbank.org/en/news/press-release/2021/03/18/nepal-receives-75-million-for-covid-19-vaccines-and-stronger-response-to-pandemic}.

\bibitem[BBC(2021{\natexlab{a}})]{Nw6}
BBC.
\newblock Covid passports: What are different countries planning?,
  2021{\natexlab{a}}.
\newblock URL \url{https://www.bbc.com/news/world-europe-56522408}.

\bibitem[BBC(2021{\natexlab{b}})]{SKorea5}
BBC.
\newblock South korea coronavirus: Pm aims for 'herd immunity by autumn',
  2021{\natexlab{b}}.
\newblock URL \url{https://www.bbc.com/news/world-asia-56156234}.

\bibitem[BBC(2021{\natexlab{c}})]{india7}
BBC.
\newblock Coronavirus: 'double mutant' covid variant found in india,
  2021{\natexlab{c}}.
\newblock URL \url{https://www.bbc.com/news/world-asia-india-56507988}.

\bibitem[BeijingReview(2020)]{china10}
BeijingReview.
\newblock China's covid vaccines to be produced, distributed under strict
  management, 2020.
\newblock URL
  \url{http://www.bjreview.com/Nation/202012/t20201231_800231511.html}.

\bibitem[Bernama(2021)]{Malaysia15}
Bernama.
\newblock Dr adham: Registration for covid-19 vaccination to open soon, 2021.
\newblock URL
  \url{https://www.nst.com.my/news/nation/2021/01/657602/dr-adham-registration-covid-19-vaccination-open-soon}.

\bibitem[Bloomberg(2020)]{Japan2}
Bloomberg.
\newblock A bitter vaccine history means hurdles for japan’s covid fight,
  2020.
\newblock URL
  \url{https://www.bloomberg.com/news/articles/2020-12-22/a-bitter-vaccine-history-means-hurdles-for-japan-s-covid-fight}.

\bibitem[Bloomberg(2021)]{Malaysia21}
Bloomberg.
\newblock Malaysia recovery seen driven by tech upcycle, vaccines, 2021.
\newblock URL
  \url{https://www.bloomberg.com/news/articles/2021-04-26/malaysia-recovery-to-be-driven-by-tech-upcycle-vaccines-jobs}.

\bibitem[Board(2021)]{Estonia6}
Health Board.
\newblock Republic of estonia, 2021.
\newblock URL \url{https://terviseamet.ee/en}.

\bibitem[Business(2021)]{HK7}
Hong~Kong Business.
\newblock Hong kong budget 2021: \$8.4b for covid-19 vaccines; no new taxes,
  2021.
\newblock URL
  \url{https://hongkongbusiness.hk/economy/news/hong-kong-budget-2021-84b-covid-19-vaccines-no-new-taxes}.

\bibitem[BusinessToday(2021)]{china5}
BusinessToday.
\newblock China begins covid-19 vaccination drive; over 73,000 receive first
  shot, 2021.
\newblock URL
  \url{https://www.businesstoday.in/current/economy-politics/china-begins-covid-19-vaccination-drive-over-73000-receive-first\\-shot/story/426896.html}.

\bibitem[CABAR(2021)]{Kaza1}
CABAR.
\newblock What should kazakhstan citizens know about the upcoming covid-19
  vaccination?, 2021.
\newblock URL
  \url{https://cabar.asia/en/what-should-kazakhstan-citizens-know-about-the-upcoming-covid-19-vaccination}.

\bibitem[Canada(2021)]{canada5}
Statistics Canada.
\newblock Canadian international merchandise trade, january 2021, 2021.
\newblock URL
  \url{https://www150.statcan.gc.ca/n1/daily-quotidien/210305/dq210305a-eng.htm}.

\bibitem[ChinaDaily(2021)]{china8}
ChinaDaily.
\newblock Covid-19 vaccine: Answers to frequently asked questions, 2021.
\newblock URL \url{http://en.nhc.gov.cn/2021-01/08/c_82699.htm}.

\bibitem[CNA(2020{\natexlab{a}})]{Malaysia1}
CNA.
\newblock Malaysia pm muhyiddin to be among the first in country to take
  covid-19 vaccine, 2020{\natexlab{a}}.
\newblock URL
  \url{https://www.channelnewsasia.com/news/asia/malaysia-pm-muhyiddin-yassin-first-take-covid-19-vaccine-13822286}.

\bibitem[CNA(2020{\natexlab{b}})]{Malaysia9}
CNA.
\newblock Malaysia agrees to buy 12.8 million doses of pfizer's covid-19
  vaccine, 2020{\natexlab{b}}.
\newblock URL
  \url{https://www.channelnewsasia.com/news/asia/malaysia-covid-19-pfizer-vaccine-coronavirus-13651672}.

\bibitem[CNBC(2020)]{Malaysia10}
CNBC.
\newblock Malaysia buys astrazeneca covid-19 vaccines, seeks more from china,
  russia, 2020.
\newblock URL
  \url{https://www.cnbc.com/2020/12/22/malaysia-procures-6point4-million-doses-astrazeneca-coronavirus-vaccine.html}.

\bibitem[CNBC(2021{\natexlab{a}})]{Russ7}
CNBC.
\newblock Putin gets a covid vaccine — but the kremlin refuses to say which
  one, 2021{\natexlab{a}}.
\newblock URL
  \url{https://www.cnbc.com/2021/03/23/putin-to-get-coronavirus-vaccine-russias-vaccine-strategy-in-focus.html}.

\bibitem[CNBC(2021{\natexlab{b}})]{Singapore12}
CNBC.
\newblock Singapore sets aside over \$8 billion in 2021 budget for new covid
  support package, 2021{\natexlab{b}}.
\newblock URL
  \url{https://www.cnbc.com/2021/02/16/singapore-announces-2021-government-budget-to-support-covid-recovery.html}.

\bibitem[CodeBlue(2021)]{Malaysia16}
CodeBlue.
\newblock Mysejahtera for covid vaccine registration, reporting side effects:
  Khairy, 2021.
\newblock URL
  \url{https://codeblue.galencentre.org/2021/01/06/mysejahtera-for-covid-vaccine-registration-reporting-side-effects-khairy/}.

\bibitem[Costruiresalute.it(2020)]{It5}
Costruiresalute.it.
\newblock Vaccini, 2020.
\newblock URL \url{https://testcovid.costruiresalute.it/}.

\bibitem[della Salute(2021)]{It1}
Ministero della Salute.
\newblock Piano vaccini anti covid-19, 2021.
\newblock URL
  \url{http://www.salute.gov.it/imgs/C_17_pubblicazioni_2986_allegato.pdf}.

\bibitem[della Sera(2020)]{It6}
Corriere della Sera.
\newblock Vaccino covid, arcuri: campagna da metà gennaio in padiglioni a
  forma di primula disegnati da stefano boeri, 2020.
\newblock URL
  \url{https://www.corriere.it/cronache/20_dicembre_13/vaccino-covid-arcuri-campagna-meta-gennaio-padiglioni-forma-primula-disegnati-stefano-boeri-1478a0b6-3d2e-11eb-943e-95a1c9e91e01.shtml}.

\bibitem[des Solidarités et de~la Santé(2020)]{F2}
Ministère des Solidarités et de~la Santé.
\newblock La stratégie vaccinale et la liste des publics prioritaires, 2020.
\newblock URL
  \url{https://solidarites-sante.gouv.fr/grands-dossiers/vaccin-covid-19/publics-prioritaires-vaccin-covid-19#liste-pro}.

\bibitem[des Solidarités et de~la Santé(2021)]{F1}
Ministère des Solidarités et de~la Santé.
\newblock Trouver un lieu de vaccination covid-19, 2021.
\newblock URL \url{https://www.sante.fr/cf/centres-vaccination-covid.html}.

\bibitem[di~Sanità(2020)]{It3}
Istituto~Superiore di~Sanità.
\newblock Piano nazionale di vaccinazione covid-19, 2020.
\newblock URL
  \url{https://www.epicentro.iss.it/vaccini/covid-19-piano-vaccinazione}.

\bibitem[Egov(2021)]{Kaza5}
Egov.
\newblock Vaccination against the coronavirus infection, 2021.
\newblock URL
  \url{https://egov.kz/cms/kk/articles/health_care/Vakcinaciya-protiv-koronavirusnoy-infekcii-}.

\bibitem[Ella et~al.(2020)Ella, Reddy, Jogdand, Sarangi, Ganneru, Prasad, Das,
  Raju, Praturi, Sapkal, et~al.]{ella2020safety}
Raches Ella, Siddharth Reddy, Harsh Jogdand, Vamshi Sarangi, Brunda Ganneru,
  Sai Prasad, Dipankar Das, Dugyala Raju, Usha Praturi, Gajanan Sapkal, et~al.
\newblock Safety and immunogenicity clinical trial of an inactivated sars-cov-2
  vaccine, bbv152 (a phase 2, double-blind, randomised controlled trial) and
  the persistence of immune responses from a phase 1 follow-up report.
\newblock \emph{medRxiv}, 2020.

\bibitem[EuroNews(2020)]{It11}
EuroNews.
\newblock Italy's 2021 budget attempts to cushion covid-19 financial impact,
  2020.
\newblock URL
  \url{https://www.euronews.com/2020/11/18/italy-s-2021-budget-attempts-to-cushion-covid-19-financial-impact}.

\bibitem[Euronews(2021)]{Russ8}
Euronews.
\newblock The pros and cons of vaccine passports: Russian foreign minister
  latest to voice concerns, 2021.
\newblock URL
  \url{https://www.euronews.com/travel/2021/03/23/the-pros-and-cons-of-vaccine-passports-russian-foreign-minister-latest-to-voice-concerns}.

\bibitem[EuropeanPharmaceuticalReview(2021)]{Japan16}
EuropeanPharmaceuticalReview.
\newblock Japan has pre-ordered covid-19 vaccine doses for more than four times
  its population, 2021.
\newblock URL
  \url{https://www.europeanpharmaceuticalreview.com/news/140133/japan-has-pre-ordered-covid-19-vaccine-doses-for-more-than-four\\-times-its-population-says-report/}.

\bibitem[Express(2021)]{canada7}
The~Indian Express.
\newblock Canadian researchers release first image of b.1.1.7 variant of
  covid-19, exude faith in existing vaccines, 2021.
\newblock URL
  \url{https://indianexpress.com/article/coronavirus/ubc-researches-release-first-image-b-1-1-7-variant-covid-19-faith-vaccines-7301544/}.

\bibitem[Folkehelseinstituttet(2020)]{Nw2}
Folkehelseinstituttet.
\newblock Covid-19 vaksinasjonsprogrammet, 2020.
\newblock URL
  \url{https://www.fhi.no/contentassets/1af4c6e655014a738055c79b72396de8/vedlegg/mulige-kriterier-for-a-prioritere-mellom-helsepersonellgrupper-i-spesialisthelsetjenesten_.pdf}.

\bibitem[for Communicable~Diseases(2021)]{SA3}
National~Institute for Communicable~Diseases.
\newblock Covid-19 vaccine rollout strategy faq, 2021.
\newblock URL
  \url{https://www.nicd.ac.za/covid-19-vaccine-rollout-strategy-faq/}.

\bibitem[for Ensuring Access~to COVID-19 Vaccine
  Supply~(JKJAV)(2021{\natexlab{a}})]{Gm0}
The Special~Committee for Ensuring Access~to COVID-19 Vaccine Supply~(JKJAV).
\newblock Explained: When will i be in line for a covid-19 vaccination in
  germany?, 2021{\natexlab{a}}.
\newblock URL
  \url{https://www.thelocal.de/20210309/explained-when-will-i-be-in-line-for-a-vaccination-in-germany/}.

\bibitem[for Ensuring Access~to COVID-19 Vaccine
  Supply~(JKJAV)(2021{\natexlab{b}})]{Malaysia20}
The Special~Committee for Ensuring Access~to COVID-19 Vaccine Supply~(JKJAV).
\newblock National covid-19 immunisation programme. location of vaccination
  centres, 2021{\natexlab{b}}.
\newblock URL
  \url{https://www.vaksincovid.gov.my/pdf/National_COVID-19_Immunisation_Programme_English.pdf#page=21}.

\bibitem[for Systems~Science \& Engineering(2021)for Systems~Science and
  Engineering]{Nepal7}
Johns Hopkins University~Center for Systems~Science and Engineering.
\newblock Covid-19 data repository by the center for systems science and
  engineering (csse) at johns hopkins university, 2021.
\newblock URL \url{https://github.com/CSSEGISandData/COVID-19}.

\bibitem[France24(2020)]{F5}
France24.
\newblock Experts welcome france’s covid-19 vaccine strategy – with
  caveats, 2020.
\newblock URL
  \url{https://www.france24.com/en/europe/20201208-experts-welcome-france-s-covid-19-vaccine-strategy-–-with-caveats}.

\bibitem[Gavi(2021)]{Kaza4}
Gavi.
\newblock Covax vaccine roll-out, 2021.
\newblock URL \url{https://www.gavi.org/covax-vaccine-roll-out}.

\bibitem[Ghosh(2021)]{india4}
Abantika Ghosh.
\newblock What is cowin and what you need to register on the app for covid
  vaccine shot, 2021.
\newblock URL
  \url{https://theprint.in/health/what-is-cowin-and-what-you-need-to-register-on-the-app-for-cov\\id-vaccine-shot/579307/}.

\bibitem[GlobalTimes(2020)]{china9}
GlobalTimes.
\newblock China joins covax, to provide vaccines to developing countries first,
  2020.
\newblock URL \url{https://www.globaltimes.cn/content/1202940.shtml}.

\bibitem[Government(2021{\natexlab{a}})]{Estonia3}
Estonian Government.
\newblock Frequently asked questions, 2021{\natexlab{a}}.
\newblock URL \url{https://www.kriis.ee/en/korduma-kippuvad-kysimused}.

\bibitem[Government(2021{\natexlab{b}})]{HK4}
Hong Kong Special Administrative~Region Government.
\newblock Covid-19 vaccination programme, 2021{\natexlab{b}}.
\newblock URL \url{https://www.covidvaccine.gov.hk/en/programme}.

\bibitem[Government(2021{\natexlab{c}})]{HK6}
Hong Kong Special Administrative~Region Government.
\newblock About the vaccines, 2021{\natexlab{c}}.
\newblock URL \url{https://www.covidvaccine.gov.hk/en/vaccine}.

\bibitem[GovtofCanada(2021{\natexlab{a}})]{canada1}
GovtofCanada.
\newblock Covid-19 vaccine: Guidance on the prioritization of initial doses,
  2021{\natexlab{a}}.
\newblock URL
  \url{https://www.canada.ca/en/public-health/services/immunization/national-advisory-committee-on-immunization-naci/guidance-prioritization-initial-doses-covid-19-vaccines.html}.

\bibitem[GovtofCanada(2021{\natexlab{b}})]{canada2}
GovtofCanada.
\newblock Covid-19: Preliminary guidance on key populations for early
  immunization, 2021{\natexlab{b}}.
\newblock URL
  \url{https://www.canada.ca/en/public-health/services/immunization/national-advisory-committee-on-immunization-naci/guidance-key-populations-early-covid-19-immunization.html}.

\bibitem[GovtofCanada(2021{\natexlab{c}})]{canada3}
GovtofCanada.
\newblock Covid-19 immunization plan, 2021{\natexlab{c}}.
\newblock URL
  \url{https://www2.gov.bc.ca/gov/content/safety/emergency-preparedness-response-recovery/covid-19-provincial-support/vaccines}.

\bibitem[GovtofCanada(2021{\natexlab{d}})]{canada4}
GovtofCanada.
\newblock Procuring vaccines for covid-19, 2021{\natexlab{d}}.
\newblock URL
  \url{https://www.canada.ca/en/public-services-procurement/services/procuring-vaccines-covid19.html}.

\bibitem[GovtofIndia(2020)]{india5}
GovtofIndia.
\newblock Covid-19 vaccines operational guidelines, 2020.
\newblock URL
  \url{https://www.mohfw.gov.in/pdf/COVID19VaccineOG111Chapter16.pdf}.

\bibitem[GovtofTurkey(2021{\natexlab{a}})]{Turkey12}
GovtofTurkey.
\newblock Covid-19 aşısı kayıt ve bildirim, 2021{\natexlab{a}}.
\newblock URL
  \url{https://covid19asi.saglik.gov.tr/TR-77821/covid-19-asisi-kayit-ve-bildirim.html}.

\bibitem[GovtofTurkey(2021{\natexlab{b}})]{Turkey13}
GovtofTurkey.
\newblock Aşı takip sistemi, 2021{\natexlab{b}}.
\newblock URL
  \url{https://covid19asi.saglik.gov.tr/TR-77807/asi-takip-sistemi.html}.

\bibitem[Guardtime(2021)]{Estonia7}
Guardtime.
\newblock Vaccineguard, 2021.
\newblock URL \url{https://guardtime.com/vaccineguard}.

\bibitem[Hankyoreh(2021)]{SKorea8}
The Hankyoreh.
\newblock 1 in 4 new covid-19 cases in s. korea are variant cases, 2021.
\newblock URL
  \url{http://english.hani.co.kr/arti/english_edition/e_national/994909.html}.

\bibitem[Health \& Centre(2021)Health and Centre]{Estonia5}
Health and Welfare Information~Systems Centre.
\newblock Patient portal, 2021.
\newblock URL \url{https://www.digilugu.ee/login?locale=en}.

\bibitem[Healthworld(2021)]{SKorea2}
Healthworld.
\newblock South korea to vaccine its 52 million people for free, 2021.
\newblock URL
  \url{https://health.economictimes.indiatimes.com/news/industry/south-korea-to-vaccine-its-52-million-people-for-free/80213666}.

\bibitem[Hindu(2021)]{Nepal9}
The Hindu.
\newblock Nepal runs out of vaccine, waiting for shipment from india, 2021.
\newblock URL
  \url{https://www.thehindu.com/news/national/nepal-runs-out-of-vaccine-waiting-for-shipment-from-india/article34275648.ece}.

\bibitem[Hour(2021)]{SA8}
PBS~News Hour.
\newblock South africa to spend \$712 million on mass vaccination drive, 2021.
\newblock URL
  \url{https://www.pbs.org/newshour/world/south-africa-to-spend-712-million-on-mass-vaccination-drive}.

\bibitem[in~Data(2020)]{Ourworldindata}
Our~World in~Data.
\newblock Our world in data, 2020.
\newblock URL
  \url{https://ourworldindata.org/grapher/daily-covid-vaccination-doses-per-capita}.

\bibitem[Independent(2021)]{Eg6}
Egypt Independent.
\newblock Health minister signs protocol to vaccinate 2 million egyptians for
  free, 2021.
\newblock URL
  \url{https://egyptindependent.com/health-minister-signs-protocol-to-vaccinate-2-million-egyptians-for-free/}.

\bibitem[Ismail et~al.(2010)Ismail, Langley, Harris, Warshawsky, Desai, and
  FarhangMehr]{ismail2010canada}
Shainoor~J Ismail, Joanne~M Langley, Tara~M Harris, Bryna~F Warshawsky, Shalini
  Desai, and Mahnaz FarhangMehr.
\newblock Canada's national advisory committee on immunization (naci):
  evidence-based decision-making on vaccines and immunization.
\newblock \emph{Vaccine}, 28:\penalty0 A58--A63, 2010.

\bibitem[Ismail et~al.(2020)Ismail, Hardy, Tunis, Young, Sicard, and
  Quach]{ismail2020framework}
Shainoor~J Ismail, Kendra Hardy, Matthew~C Tunis, Kelsey Young, Nadine Sicard,
  and Caroline Quach.
\newblock A framework for the systematic consideration of ethics, equity,
  feasibility, and acceptability in vaccine program recommendations.
\newblock \emph{Vaccine}, 38\penalty0 (36):\penalty0 5861--5876, 2020.

\bibitem[Jeyanathan et~al.(2020)Jeyanathan, Afkhami, Smaill, Miller, Lichty,
  and Xing]{jeyanathan2020immunological}
Mangalakumari Jeyanathan, Sam Afkhami, Fiona Smaill, Matthew~S Miller, Brian~D
  Lichty, and Zhou Xing.
\newblock Immunological considerations for covid-19 vaccine strategies.
\newblock \emph{Nature Reviews Immunology}, 20\penalty0 (10):\penalty0
  615--632, 2020.

\bibitem[Journal(2021)]{canada8}
The Wall~Street Journal.
\newblock How a canadian province contained the brazilian covid-19 variant,
  2021.
\newblock URL
  \url{https://www.wsj.com/articles/how-a-canadian-province-contained-the-brazilian-covid-19-variant-11620641317}.

\bibitem[Kim et~al.(2021)Kim, Marks, and Clemens]{kim2021looking}
Jerome~H Kim, Florian Marks, and John~D Clemens.
\newblock Looking beyond covid-19 vaccine phase 3 trials.
\newblock \emph{Nature medicine}, pp.\  1--7, 2021.

\bibitem[Knoll \& Wonodi(2021)Knoll and Wonodi]{knoll2021oxford}
Maria~Deloria Knoll and Chizoba Wonodi.
\newblock Oxford--astrazeneca covid-19 vaccine efficacy.
\newblock \emph{The Lancet}, 397\penalty0 (10269):\penalty0 72--74, 2021.

\bibitem[Kommune(2020)]{Nw3}
Elverum Kommune.
\newblock Information about coronavirus vaccine, 2020.
\newblock URL
  \url{https://www.elverum.kommune.no/om-oss/aktuelt/information-about-coronavirus-vaccine}.

\bibitem[Kwan(2021)]{Malaysia14}
Faye Kwan.
\newblock Mysejahtera to register vaccination data soon, 2021.
\newblock URL
  \url{https://www.freemalaysiatoday.com/category/nation/2021/01/25/mysejahtera-to-register-vaccination-data-soon/}.

\bibitem[Larsen(2021)]{Malaysia17}
Mette Larsen.
\newblock Covid-19 vaccine is not mandatory in malaysia, 2021.
\newblock URL
  \url{https://scandasia.com/covid-19-vaccine-is-not-mandatory-in-malaysia/}.

\bibitem[Livemint(2021{\natexlab{a}})]{Nepal5}
Livemint.
\newblock Nepal to restart covid-19 vaccinations after china donates shots,
  2021{\natexlab{a}}.
\newblock URL
  \url{https://www.livemint.com/news/world/nepal-to-restart-covid-19-vaccinations-after-china-donates-shots-11617103028081.html}.

\bibitem[Livemint(2021{\natexlab{b}})]{Nepal6}
Livemint.
\newblock Nepal to receive 100,000 more covid vaccine doses from india,
  2021{\natexlab{b}}.
\newblock URL
  \url{https://www.livemint.com/news/world/nepal-to-receive-100-000-more-covid-vaccine-doses-from-india-11616840756054.html}.

\bibitem[Madhi et~al.(2021)Madhi, Baillie, Cutland, Voysey, Koen, Fairlie,
  Padayachee, Dheda, Barnabas, Bhorat, et~al.]{madhi2021safety}
Shabir~A Madhi, Vicky Baillie, Clare~L Cutland, Merryn Voysey, Anthonet~L Koen,
  Lee Fairlie, Sherman~D Padayachee, Keertan Dheda, Shaun~L Barnabas,
  Qasim~Ebrahim Bhorat, et~al.
\newblock Safety and efficacy of the chadox1 ncov-19 (azd1222) covid-19 vaccine
  against the b. 1.351 variant in south africa.
\newblock \emph{MedRxiv}, 2021.

\bibitem[Media(2021)]{F9}
HIMSS Media.
\newblock France to trial digital vaccine passport scheme, 2021.
\newblock URL
  \url{https://www.healthcareitnews.com/news/emea/france-trial-digital-vaccine-passport-scheme}.

\bibitem[Mills \& Salisbury(2021)Mills and Salisbury]{mills2021challenges}
Melinda~C Mills and David Salisbury.
\newblock The challenges of distributing covid-19 vaccinations.
\newblock \emph{EClinicalMedicine}, 31, 2021.

\bibitem[Mint(2021)]{Singapore13}
Mint.
\newblock New covid strain in singapore: What we know so far about b.1.617
  variant affecting children, 2021.
\newblock URL
  \url{https://www.livemint.com/science/health/new-covid-strain-in-singapore-what-we-know-so-far-about-b-1-617-variant-affecting-children-11621358293334.html}.

\bibitem[News(2021{\natexlab{a}})]{Eg8}
Arab News.
\newblock Covid-19: Egypt identifies next 10 days as dangerous for spread,
  urges caution, 2021{\natexlab{a}}.
\newblock URL \url{https://www.arabnews.com/node/1853541/middle-east}.

\bibitem[News(2021{\natexlab{b}})]{canada6}
CTV News.
\newblock Canada spent \$24m on covid-19 vaccines received in january: Statcan,
  2021{\natexlab{b}}.
\newblock URL
  \url{https://www.ctvnews.ca/health/coronavirus/canada-spent-24m-on-covid-19-vaccines-received-in-january-statcan-1.5335083}.

\bibitem[News(2021{\natexlab{c}})]{Estonia10}
ERR News.
\newblock Lanno: Rising infection rate has moved from north to south estonia,
  2021{\natexlab{c}}.
\newblock URL
  \url{https://news.err.ee/1608202870/lanno-rising-infection-rate-has-moved-from-north-to-south-estonia}.

\bibitem[News(2021{\natexlab{d}})]{Estonia9}
ERR News.
\newblock Estonia looking to donate 800,000 coronavirus vaccines,
  2021{\natexlab{d}}.
\newblock URL
  \url{https://news.err.ee/1608233595/estonia-looking-to-donate-800-000-coronavirus-vaccines}.

\bibitem[News(2021{\natexlab{e}})]{Japan18}
Kyodo News.
\newblock Japan speeds up covid vaccine rollout for elderly amid virus surge,
  2021{\natexlab{e}}.
\newblock URL
  \url{https://english.kyodonews.net/news/2021/05/d64017a7b6ea-japan-speeds-up-covid-vaccine-rollout-for-elderly-amid-virus-surge.html}.

\bibitem[News(2021{\natexlab{f}})]{Japan20}
Kyodo News.
\newblock Japan speeds up covid vaccine rollout for elderly amid virus surge,
  2021{\natexlab{f}}.
\newblock URL
  \url{https://english.kyodonews.net/news/2021/05/d64017a7b6ea-japan-speeds-up-covid-vaccine-rollout-for-elderly-amid-virus-surge.html}.

\bibitem[News(2021{\natexlab{g}})]{SA9}
VOA News.
\newblock Two highly contagious coronavirus variants appear in south africa,
  2021{\natexlab{g}}.
\newblock URL
  \url{https://www.voanews.com/covid-19-pandemic/two-highly-contagious-coronavirus-variants-appear-south-africa}.

\bibitem[News-Medical.Net(2020)]{It9}
News-Medical.Net.
\newblock New n501 sars-cov-2 mutation may have been circulating in italy since
  august 2020, 2020.
\newblock URL
  \url{https://www.news-medical.net/news/20210113/New-N501-SARS-CoV-2-mutation-may-have-been-circulating-in-Italy-since-August-2020.aspx}.

\bibitem[News24(2021)]{SA7}
News24.
\newblock Treasury to spend r19.3bn buying covid-19 vaccines, 2021.
\newblock URL
  \url{https://www.news24.com/fin24/economy/treasury-to-spend-r193bn-buying-covid-19-vaccines-20210224}.

\bibitem[Nikkei(2021)]{TH9}
Asia Nikkei.
\newblock Thailand, vietnam and cambodia hit covid variants with iron fist,
  2021.
\newblock URL
  \url{https://asia.nikkei.com/Spotlight/Coronavirus/Thailand-Vietnam-and-Cambodia-hit-COVID-variants-with-iron-fist}.

\bibitem[Norway(2021)]{Nw8}
The~Local Norway.
\newblock Several cases of new covid-19 variant detected in norway, 2021.
\newblock URL
  \url{https://www.thelocal.no/20210218/several-cases-of-new-covid-19-variant-detected-in-norway/}.

\bibitem[of~Commerce(2021)]{TH2}
Danish-Thai~Chamber of~Commerce.
\newblock Covid vaccine distribution plans for thailand, 2021.
\newblock URL
  \url{https://dancham.or.th/covid-vaccine-distribution-plans-for-thailand/}.

\bibitem[of~Health(2021{\natexlab{a}})]{VT2}
Ministry of~Health.
\newblock Need to know: Guidance on storage, distribution and use of covid-19
  vaccines, 2021{\natexlab{a}}.
\newblock URL \url{https://ncov.moh.gov.vn/vi/web/guest/-/6847426-1892}.

\bibitem[of~Health(2021{\natexlab{b}})]{VT3}
Ministry of~Health.
\newblock Ministry of health distributes vaccine covid-19 phase 2,
  2021{\natexlab{b}}.
\newblock URL \url{https://ncov.moh.gov.vn/vi/web/guest/-/6851640-33}.

\bibitem[of~Health(2021{\natexlab{c}})]{VT4}
Ministry of~Health.
\newblock The government added 502.9 billion vnd for covid-19 prevention and
  control, 2021{\natexlab{c}}.
\newblock URL \url{https://ncov.moh.gov.vn/vi/web/guest/-/6851640-30}.

\bibitem[of~India(2020)]{china13}
The~Times of~India.
\newblock Chinese covid-19 vaccine maker gets \$500 million funding boost,
  2020.
\newblock URL
  \url{https://timesofindia.indiatimes.com/world/china/chinese-covid-19-vaccine-maker-gets-500-million-funding-boost/articleshow/79606524.cms}.

\bibitem[of~India(2021)]{VT8}
Times of~India.
\newblock New vietnam coronavirus variant: What we know so far and why we need
  to be careful, 2021.
\newblock URL
  \url{https://timesofindia.indiatimes.com/life-style/health-fitness/health-news/new-vietnam-coronavirus-covid-variant-what-we-know-so-far-and-why-we-need-to-be-careful/articleshow/83096223.cms}.

\bibitem[of~Malaysia Official~Website(2020)]{Malaysia13}
Prime Minister's~Office of~Malaysia Official~Website.
\newblock Covid-19 vaccine to be given free to malaysians – pm muhyiddin,
  2020.
\newblock URL
  \url{https://www.pmo.gov.my/2020/11/covid-19-vaccine-to-be-given-free-to-malaysians-pm-muhyiddin/}.

\bibitem[of~Public~Health(2021{\natexlab{a}})]{Nw1}
Norwegian~Institute of~Public~Health.
\newblock Development and approval of coronavirus vaccine, 2021{\natexlab{a}}.
\newblock URL
  \url{https://www.fhi.no/en/id/vaccines/coronavirus-immunisation-programme/development-of-covid-19-vaccine/}.

\bibitem[of~Public~Health(2021{\natexlab{b}})]{Nw4}
Norwegian~Institute of~Public~Health.
\newblock Astrazeneca vaccine still on hold in norway, 2021{\natexlab{b}}.
\newblock URL
  \url{https://www.fhi.no/en/news/2021/fortsatt-pause-for-astrazeneca-vaksinen-i-norge/}.

\bibitem[of~Public~Health(2021{\natexlab{c}})]{Nw5}
Norwegian~Institute of~Public~Health.
\newblock Norwegian institute of public health, 2021{\natexlab{c}}.
\newblock URL \url{https://www.fhi.no/en/}.

\bibitem[of~Public~Health(2021{\natexlab{d}})]{Nw7}
Norwegian~Institute of~Public~Health.
\newblock Norwegian institute of public health's recommendation about
  astrazeneca vaccine, 2021{\natexlab{d}}.
\newblock URL
  \url{https://www.fhi.no/en/news/2021/astrazeneca-vaccine-removed-from-coronavirus-immunisation-programme-in-norw/}.

\bibitem[of~Sciences~Engineering et~al.(2020)of~Sciences~Engineering, Medicine,
  et~al.]{national2020framework}
National~Academies of~Sciences~Engineering, Medicine, et~al.
\newblock Framework for equitable allocation of covid-19 vaccine.
\newblock \emph{The National Academies Press}, 2020.

\bibitem[of~Social~Affairs(2021)]{Estonia1}
Ministry of~Social~Affairs.
\newblock Home site, 2021.
\newblock URL \url{https://www.sm.ee/en}.

\bibitem[of~South~Africa(2021)]{SA1}
Government of~South~Africa.
\newblock Covid-19 coronavirus vaccine strategy, 2021.
\newblock URL \url{https://www.gov.za/covid-19/vaccine/strategy}.

\bibitem[of~Thailand(2021)]{TH1}
The Government Public Relations~Department of~Thailand.
\newblock The government public relations department, 2021.
\newblock URL \url{https://thailand.prd.go.th/ewt_news.php?nid=10847}.

\bibitem[of~Turkey(2021)]{Turkey11}
Govt of~Turkey.
\newblock Covid-19 aşısı temin ve dağıtım süreci, 2021.
\newblock URL
  \url{https://covid19asi.saglik.gov.tr/TR-77822/covid-19-asisi-temin-ve-dagitim-sureci.html}.

\bibitem[Online(2020)]{Malaysia4}
The~Star Online.
\newblock Pharmaniaga to distribute covid-19 vaccines efficiently, prioritise
  safety aspects, says group md, 2020.
\newblock URL
  \url{https://www.thestar.com.my/business/business-news/2020/12/28/pharmaniaga-set-to-distribute-vaccines-efficiently-prioritising-safety-aspects}.

\bibitem[Opengovasia(2020)]{SKorea4}
Opengovasia.
\newblock South korea looks to tech to combat covid-19, 2020.
\newblock URL
  \url{https://opengovasia.com/south-korea-looks-to-tech-to-combat-covid-19/}.

\bibitem[opengovasia.com(2020)]{VT6}
opengovasia.com.
\newblock Vietnam launches health app to manage covid-19, 2020.
\newblock URL
  \url{https://opengovasia.com/vietnam-launches-health-app-to-manage-covid-19/}.

\bibitem[Organization(2021{\natexlab{a}})]{Eg3}
World~Health Organization.
\newblock 854 400 covid-19 vaccine doses shipped by covax facility to egypt,
  2021{\natexlab{a}}.
\newblock URL
  \url{https://www.emro.who.int/media/news/854-400-covid-19-vaccine-doses-shipped-by-covax-to-egypt.html}.

\bibitem[Organization(2021{\natexlab{b}})]{Nepal8}
World~Health Organization.
\newblock Covax reaches over 100 economies, 42 days after first international
  delivery, 2021{\natexlab{b}}.
\newblock URL
  \url{https://www.who.int/news/item/08-04-2021-covax-reaches-over-100-economies-42-days-after-first-international-delivery}.

\bibitem[Organization(2021{\natexlab{c}})]{SA4}
World~Health Organization.
\newblock 172 countries and multiple candidate vaccines engaged in covid-19
  vaccine global access facility, 2021{\natexlab{c}}.
\newblock URL
  \url{https://www.who.int/news/item/24-08-2020-172-countries-and-multiple-candidate-vaccines-engaged-in-covid-19-vaccine-global-access-facility}.

\bibitem[Persad et~al.(2020)Persad, Peek, and Emanuel]{persad2020fairly}
Govind Persad, Monica~E Peek, and Ezekiel~J Emanuel.
\newblock Fairly prioritizing groups for access to covid-19 vaccines.
\newblock \emph{Jama}, 324\penalty0 (16):\penalty0 1601--1602, 2020.

\bibitem[Pfordten(2021)]{Malaysia12}
Diyana Pfordten.
\newblock Interactive: When to expect your covid-19 vaccine shots, and
  everything else you need to know about malaysia’s vaccination programme,
  2021.
\newblock URL
  \url{https://www.thestar.com.my/news/nation/2021/01/30/interactive-when-to-expect-your-covid-19-vaccine-shots-and-everything-else-you-need-to-know-about-malaysias-vaccination-programme}.

\bibitem[Piraveenan et~al.(2020)Piraveenan, Sawleshwarkar, Walsh, Zablotska,
  Bhattacharyya, Farooqui, Bhatnagar, Karan, Murhekar, Zodpey,
  et~al.]{piraveenan2020optimal}
Mahendra Piraveenan, Shailendra Sawleshwarkar, Michael Walsh, Iryna Zablotska,
  Samit Bhattacharyya, Habib~Hassan Farooqui, Tarun Bhatnagar, Anup Karan,
  Manoj Murhekar, Sanjay Zodpey, et~al.
\newblock Optimal governance and implementation of vaccination programs to
  contain the covid-19 pandemic.
\newblock \emph{arXiv preprint arXiv:2011.06455}, 2020.

\bibitem[Polack et~al.(2020)Polack, Thomas, Kitchin, Absalon, Gurtman,
  Lockhart, Perez, P{\'e}rez~Marc, Moreira, Zerbini, et~al.]{polack2020safety}
Fernando~P Polack, Stephen~J Thomas, Nicholas Kitchin, Judith Absalon,
  Alejandra Gurtman, Stephen Lockhart, John~L Perez, Gonzalo P{\'e}rez~Marc,
  Edson~D Moreira, Cristiano Zerbini, et~al.
\newblock Safety and efficacy of the bnt162b2 mrna covid-19 vaccine.
\newblock \emph{New England Journal of Medicine}, 383\penalty0 (27):\penalty0
  2603--2615, 2020.

\bibitem[Portal(2021)]{SA5}
COVID-19 South African~Online Portal.
\newblock Vaccine news, updates \& information portal, 2021.
\newblock URL \url{https://www.gov.za/covid-19/vaccine/strategy}.

\bibitem[Post(2020{\natexlab{a}})]{TH7}
Bangkok Post.
\newblock B6bn covid vaccine budget, 2020{\natexlab{a}}.
\newblock URL
  \url{https://www.bangkokpost.com/thailand/general/2021231/b6bn-covid-vaccine-budget}.

\bibitem[Post(2021{\natexlab{a}})]{TH6}
Bangkok Post.
\newblock Govt to launch 'health wallet' app for covid-19 vaccine registration,
  2021{\natexlab{a}}.
\newblock URL
  \url{https://www.bangkokpost.com/thailand/general/2093951/govt-to-launch-health-wallet-app-for-covid-19-vaccine-registration}.

\bibitem[Post(2020{\natexlab{b}})]{F4}
Financial Post.
\newblock How france will distribute covid-19 vaccines, 2020{\natexlab{b}}.
\newblock URL
  \url{https://financialpost.com/pmn/business-pmn/how-france-will-distribute-covid-19-vaccines-2}.

\bibitem[Post(2021{\natexlab{b}})]{HK8}
South China~Morning Post.
\newblock Coronavirus: mutated variant from india found in 10 hong kong
  covid-19 cases, 2021{\natexlab{b}}.
\newblock URL
  \url{https://www.scmp.com/news/hong-kong/health-environment/article/3131755/coronavirus-least-10-previously-confirmed-hong}.

\bibitem[Post(2021{\natexlab{c}})]{Nepal10}
The~Kathmandu Post.
\newblock Primary health centres to administer covid-19 vaccine to expedite
  first phase, 2021{\natexlab{c}}.
\newblock URL
  \url{https://kathmandupost.com/health/2021/01/31/primary-health-centres-to-administer-covid-19-vaccine-to-expedite-first-phase}.

\bibitem[Print(2021)]{Turkey17}
The Print.
\newblock ‘double mutant, triple mutant, bengal lineage’ — covid variants
  driving india surge decoded, 2021.
\newblock URL
  \url{https://theprint.in/science/double-mutant-triple-mutant-bengal-lineage-covid-variants-driving-india-surge-decoded/643766/}.

\bibitem[Repubblica(2020)]{It2}
La~Repubblica.
\newblock Domande e risposte sul vaccino. sicurezza, efficacia ed effetti
  collaterali, le 10 cose da sapere, 2020.
\newblock URL
  \url{https://www.repubblica.it/cronaca/2020/12/28/news/domande_e_risposte_il_vaccino_ai_raggi_x_e_l_italia_covid-free_speranza_d_autunno-280245263/}.

\bibitem[Reuters(2020{\natexlab{a}})]{F11}
Reuters.
\newblock France says covid-19 vaccine will be free for all,
  2020{\natexlab{a}}.
\newblock URL
  \url{https://www.reuters.com/article/us-health-coronavirus-vaccination-idUKKBN28D2PS}.

\bibitem[Reuters(2020{\natexlab{b}})]{Nw9}
Reuters.
\newblock Norway pledges \$1 billion to vaccines against covid-19, other
  diseases, 2020{\natexlab{b}}.
\newblock URL
  \url{https://www.reuters.com/article/us-health-coronavirus-eu-norway-idUSKBN22G1NZ}.

\bibitem[Reuters(2020{\natexlab{c}})]{Russ2}
Reuters.
\newblock Moscow rolls out sputnik v covid-19 vaccine to most exposed groups,
  2020{\natexlab{c}}.
\newblock URL
  \url{https://www.reuters.com/article/health-coronavirus-russia-vaccination/moscow-rolls-out-sputnik-v-covid-19-vaccine-to-most-exposed-groups-idINKBN28F09G}.

\bibitem[Reuters(2021{\natexlab{a}})]{Eg1}
Reuters.
\newblock Egypt begins vaccine rollout to wider population, 2021{\natexlab{a}}.
\newblock URL
  \url{https://www.reuters.com/article/uk-health-coronavirus-egypt-vaccine-idUSKCN2AW15U}.

\bibitem[Reuters(2021{\natexlab{b}})]{Eg9}
Reuters.
\newblock Reuters covid-19 tracker, 2021{\natexlab{b}}.
\newblock URL
  \url{https://graphics.reuters.com/world-coronavirus-tracker-and-maps/countries-and-territories/egypt/}.

\bibitem[Reuters(2021{\natexlab{c}})]{Estonia11}
Reuters.
\newblock Reuters covid-19 tracker, 2021{\natexlab{c}}.
\newblock URL
  \url{https://graphics.reuters.com/world-coronavirus-tracker-and-maps/countries-and-territories/estonia/}.

\bibitem[Reuters(2021{\natexlab{d}})]{Estonia4}
Reuters.
\newblock World coronavirus tracker and maps, 2021{\natexlab{d}}.
\newblock URL
  \url{https://graphics.reuters.com/world-coronavirus-tracker-and-maps/vaccination-rollout-and-access/}.

\bibitem[Reuters(2021{\natexlab{e}})]{F12}
Reuters.
\newblock Around 20 people in france detected with indian covid variant -health
  minister, 2021{\natexlab{e}}.
\newblock URL
  \url{https://www.reuters.com/world/europe/around-20-people-france-detected-with-indian-covid-variant-health-minister-2021-05-10/}.

\bibitem[Reuters(2021{\natexlab{f}})]{Gm14}
Reuters.
\newblock Berlin earmarks 9 billion euros to buy covid-19 shots for europeans,
  2021{\natexlab{f}}.
\newblock URL
  \url{https://www.reuters.com/article/us-health-coronavirus-germany-vaccines-idUSKBN2A91QI}.

\bibitem[Reuters(2021{\natexlab{g}})]{HK9}
Reuters.
\newblock Reuters covid-19 tracker, 2021{\natexlab{g}}.
\newblock URL
  \url{https://graphics.reuters.com/world-coronavirus-tracker-and-maps/countries-and-territories/hong-kong/}.

\bibitem[Reuters(2021{\natexlab{h}})]{It12}
Reuters.
\newblock Reuters covid-19 tracker, 2021{\natexlab{h}}.
\newblock URL
  \url{https://graphics.reuters.com/world-coronavirus-tracker-and-maps/countries-and-territories/italy/}.

\bibitem[Reuters(2021{\natexlab{i}})]{Kaza6}
Reuters.
\newblock Reuters covid-19 tracker, 2021{\natexlab{i}}.
\newblock URL
  \url{https://graphics.reuters.com/world-coronavirus-tracker-and-maps/countries-and-territories/kazakhstan/}.

\bibitem[Reuters(2021{\natexlab{j}})]{Malaysia22}
Reuters.
\newblock Malaysia reports first case of indian covid-19 variant,
  2021{\natexlab{j}}.
\newblock URL
  \url{https://www.reuters.com/world/india/malaysia-reports-first-case-indian-covid-19-variant-2021-05-02/}.

\bibitem[Reuters(2021{\natexlab{k}})]{Nepal15}
Reuters.
\newblock Reuters covid-19 tracker, 2021{\natexlab{k}}.
\newblock URL
  \url{https://graphics.reuters.com/world-coronavirus-tracker-and-maps/countries-and-territories/nepal/}.

\bibitem[Reuters(2021{\natexlab{l}})]{Nw10}
Reuters.
\newblock Reuters covid-19 tracker, 2021{\natexlab{l}}.
\newblock URL
  \url{https://graphics.reuters.com/world-coronavirus-tracker-and-maps/countries-and-territories/norway/}.

\bibitem[Reuters(2021{\natexlab{m}})]{Russ12}
Reuters.
\newblock Russian watchdog denies report on first cases of indian covid
  variant, 2021{\natexlab{m}}.
\newblock URL
  \url{https://www.reuters.com/world/russia-records-first-cases-indian-covid-variant-kommersant-2021-05-13/}.

\bibitem[Reuters(2021{\natexlab{n}})]{Russ13}
Reuters.
\newblock Reuters covid-19 tracker, 2021{\natexlab{n}}.
\newblock URL
  \url{https://graphics.reuters.com/world-coronavirus-tracker-and-maps/countries-and-territories/russia/}.

\bibitem[Reuters(2021{\natexlab{o}})]{Russ5}
Reuters.
\newblock Russia approves its third covid-19 vaccine, covivac,
  2021{\natexlab{o}}.
\newblock URL
  \url{https://www.reuters.com/article/us-health-coronavirus-russia-vaccine/russia-approves-its-third-covid-19-vaccine-covivac-idUSKBN2AK07H}.

\bibitem[Reuters(2021{\natexlab{p}})]{SA10}
Reuters.
\newblock Reuters covid-19 tracker, 2021{\natexlab{p}}.
\newblock URL
  \url{https://graphics.reuters.com/world-coronavirus-tracker-and-maps/countries-and-territories/south-africa/}.

\bibitem[Reuters(2021{\natexlab{q}})]{SKorea3}
Reuters.
\newblock South korea to issue blockchain-protected digital 'vaccine
  passports', 2021{\natexlab{q}}.
\newblock URL
  \url{https://www.reuters.com/article/us-health-coronavirus-southkorea-idUSKBN2BO43W}.

\bibitem[Reuters(2021{\natexlab{r}})]{SKorea6}
Reuters.
\newblock S.korea says 81\% of elderly signed up for covid-19 vaccination,
  2021{\natexlab{r}}.
\newblock URL
  \url{https://news.yahoo.com/korea-says-81-elderly-signed-084723659.html}.

\bibitem[Reuters(2021{\natexlab{s}})]{SKorea9}
Reuters.
\newblock Reuters covid-19 tracker, 2021{\natexlab{s}}.
\newblock URL
  \url{https://graphics.reuters.com/world-coronavirus-tracker-and-maps/countries-and-territories/south-korea/}.

\bibitem[Reuters(2021{\natexlab{t}})]{Singapore14}
Reuters.
\newblock Reuters covid-19 tracker, 2021{\natexlab{t}}.
\newblock URL
  \url{https://graphics.reuters.com/world-coronavirus-tracker-and-maps/countries-and-territories/singapore/}.

\bibitem[Reuters(2021{\natexlab{u}})]{TH10}
Reuters.
\newblock Reuters covid-19 tracker, 2021{\natexlab{u}}.
\newblock URL
  \url{https://graphics.reuters.com/world-coronavirus-tracker-and-maps/countries-and-territories/thailand/}.

\bibitem[Reuters(2021{\natexlab{v}})]{TH3}
Reuters.
\newblock Thailand defends decision not to join covax vaccine alliance,
  2021{\natexlab{v}}.
\newblock URL
  \url{https://www.reuters.com/article/us-health-coronavirus-thailand-idUSKBN2AE0CZ}.

\bibitem[Reuters(2021{\natexlab{w}})]{TH8}
Reuters.
\newblock B6bn covid vaccine budget, 2021{\natexlab{w}}.
\newblock URL
  \url{https://www.reuters.com/world/asia-pacific/thailand-sees-first-local-cases-indian-covid-19-variant-2021-05-21}.

\bibitem[Reuters(2021{\natexlab{x}})]{Turkey15}
Reuters.
\newblock Turkey to procure 105 million doses of covid-19 vaccines by
  end-april: Sabah, 2021{\natexlab{x}}.
\newblock URL
  \url{https://news.yahoo.com/turkey-procure-105-million-doses-142748082.html}.

\bibitem[Reuters(2021{\natexlab{y}})]{VT7}
Reuters.
\newblock Vietnam to set up \$1.1 bln covid-19 vaccine fund,
  2021{\natexlab{y}}.
\newblock URL
  \url{https://www.reuters.com/world/asia-pacific/vietnam-set-up-11-bln-covid-19-vaccine-fund-2021-05-20}.

\bibitem[Reuters(2021{\natexlab{z}})]{VT9}
Reuters.
\newblock Reuters covid-19 tracker, 2021{\natexlab{z}}.
\newblock URL
  \url{https://graphics.reuters.com/world-coronavirus-tracker-and-maps/countries-and-territories/vietnam/}.

\bibitem[Sah et~al.(2021)Sah, Khatiwada, Shrestha, Bhuvan, Tiwari, Mohapatra,
  Dhama, and Rodriguez-Morales]{Nepal2}
Ranjit Sah, Asmita~Priyadarshini Khatiwada, Sunil Shrestha, K.C. Bhuvan, Ruchi
  Tiwari, Ranjan~K. Mohapatra, Kuldeep Dhama, and Alfonso~J. Rodriguez-Morales.
\newblock Covid-19 vaccination campaign in nepal, emerging uk variant and
  futuristic vaccination strategies to combat the ongoing pandemic.
\newblock \emph{Travel Medicine and Infectious Disease}, 41:\penalty0 102037,
  2021.
\newblock ISSN 1477-8939.
\newblock \doi{https://doi.org/10.1016/j.tmaid.2021.102037}.
\newblock URL
  \url{https://www.sciencedirect.com/science/article/pii/S1477893921000788}.

\bibitem[Sahoo et~al.(2021)Sahoo, Nath, Ghosh, and Samal]{sahoo2021concepts}
Jyoti~Prakash Sahoo, Sudhanya Nath, Lipi Ghosh, and Kailash~Chandra Samal.
\newblock Concepts of immunity and recent immunization programme against
  covid-19 in india.
\newblock \emph{Biotica Research Today}, 3\penalty0 (2):\penalty0 103--106,
  2021.

\bibitem[Saieed(2021)]{Malaysia5}
Zunaira Saieed.
\newblock Pharmaniaga to supply vaccines to private sector, 2021.
\newblock URL
  \url{https://www.thestar.com.my/business/business-news/2021/02/03/pharmaniaga-to-supply-vaccines-to-private-sector}.

\bibitem[Sallam(2021)]{sallam2021covid}
Malik Sallam.
\newblock Covid-19 vaccine hesitancy worldwide: A concise systematic review of
  vaccine acceptance rates.
\newblock \emph{Vaccines}, 9\penalty0 (2):\penalty0 160, 2021.

\bibitem[Sputnikvaccine.com(2021)]{Russ4}
Sputnikvaccine.com.
\newblock Russia is the most trusted vaccine producer alongside the us, with
  sputnik v being the most recognizable vaccine, a yougov poll shows, 2021.
\newblock URL
  \url{https://sputnikvaccine.com/newsroom/pressreleases/russia-is-the-most-trusted-vaccine-producer-alongside-the-us-with-sputnik-v-being-the-most-recogniza/}.

\bibitem[Staff(2021)]{Malaysia6}
Reuters Staff.
\newblock Malaysia buys additional 12.2 million doses of pfizer-biontech
  covid-19 vaccine, 2021.
\newblock URL
  \url{https://www.reuters.com/article/us-health-coronavirus-malaysia-vaccine-idUSKBN29G0AA}.

\bibitem[Standard(2021{\natexlab{a}})]{Nepal1}
Business Standard.
\newblock Nepal gives emergency use approval to bharat biotech's covid-19
  vaccine, 2021{\natexlab{a}}.
\newblock URL
  \url{https://www.business-standard.com/article/current-affairs/nepal-gives-emergency-use-approval-to-bharat-biotech-s-covid-19-vaccine-121032000400_1.html}.

\bibitem[Standard(2021{\natexlab{b}})]{Nepal14}
Business Standard.
\newblock Nepal confirms presence of third b.1.617.2 variant of coronavirus,
  2021{\natexlab{b}}.
\newblock URL
  \url{https://www.business-standard.com/article/current-affairs/nepal-confirms-presence-of-third-b-1-617-2-variant-of-coronavirus-121051900034_1.html}.

\bibitem[statista.com(2021)]{F6}
statista.com.
\newblock Cumulative number of people who received one or two doses of vaccines
  against the coronavirus (covid-19) in france from december 27, 2020, to may
  23, 2021, 2021.
\newblock URL
  \url{https://www.statista.com/statistics/1195620/vaccines-again-covid19-france/}.

\bibitem[Streets(2021)]{Eg4}
Egyptian Streets.
\newblock What you need to know: Egypt’s vaccine distribution plan, 2021.
\newblock URL
  \url{https://egyptianstreets.com/2021/01/08/what-you-need-to-know-egypts-vaccine-distribution-plan/}.

\bibitem[Team(2021)]{SA2}
Bhekisisa Team.
\newblock Sa’s first j\&j jabs could arrive next week. here’s what our
  roll-out plan looks like, 2021.
\newblock URL
  \url{https://bhekisisa.org/health-news-south-africa/2021-02-10-what-does-sas-new-vaccine-roll-out-plan-look-like-heres-what-we-do-dont-know/}.

\bibitem[TheJapanTimes(2020)]{Japan5}
TheJapanTimes.
\newblock Reservations, gyms and thousands of freezers: Japan's vaccine rollout
  takes shape, 2020.
\newblock URL
  \url{https://www.japantimes.co.jp/news/2020/12/10/national/science-health/japan-vaccine-rollout-plans/}.

\bibitem[TheJapanTimes(2021)]{Japan14}
TheJapanTimes.
\newblock Japan to study cases of people infected after coronavirus
  vaccination, 2021.
\newblock URL
  \url{https://www.japantimes.co.jp/news/2021/01/18/national/japan-study-coronavirus-vaccination/}.

\bibitem[Times(2021{\natexlab{a}})]{It10}
Hindustan Times.
\newblock Indian covid variant detected in northern italy, 2021{\natexlab{a}}.
\newblock URL
  \url{https://www.hindustantimes.com/world-news/indian-covid-variant-detected-in-northern-italy-101619453846951.html}.

\bibitem[Times(2021{\natexlab{b}})]{Japan19}
Hindustan Times.
\newblock New coronavirus variant 'eek' found in japan, lowers vaccine
  protection, 2021{\natexlab{b}}.
\newblock URL
  \url{https://www.hindustantimes.com/world-news/new-coronavirus-variant-eek-found-in-japan-lowers-vaccine-protection-101617601060352.html}.

\bibitem[Times(2021{\natexlab{c}})]{china14}
Hindustan Times.
\newblock Covid-19: China detects 18 cases of double mutant variant from india,
  2021{\natexlab{c}}.
\newblock URL
  \url{https://www.hindustantimes.com/world-news/china-detects-18-cases-of-double-mutant-covid-19-virus-from-india-101620275569038.html}.

\bibitem[Times(2021{\natexlab{d}})]{Russ11}
The~Economic Times.
\newblock Russia records first cases of covid variant found in india,
  2021{\natexlab{d}}.
\newblock URL
  \url{https://economictimes.indiatimes.com/news/international/world-news/russia-records-first-cases-of-covid-variant-found-in-india/articleshow/82598450.cms}.

\bibitem[Times(2021{\natexlab{e}})]{india6}
The~Economic Times.
\newblock As of march 10, government spent close to rs 1,400 crore on vaccine:
  Health ministry, 2021{\natexlab{e}}.
\newblock URL
  \url{https://economictimes.indiatimes.com/news/politics-and-nation/govt-spent-close-to-1400-cr-on-vaccine/articleshow/81538018.cms}.

\bibitem[Times(2021{\natexlab{f}})]{Nepal3}
The~Himalayan Times.
\newblock Kathmandu metropolitan city designates 15 health facilities for
  covid-19 vaccination drive, 2021{\natexlab{f}}.
\newblock URL
  \url{https://thehimalayantimes.com/kathmandu/kathmandu-metropolitan-city-designates-15-health-facilities-for-covid-19-vaccination-drive}.

\bibitem[Times(2021{\natexlab{g}})]{Russ3}
The~Moscow Times.
\newblock Coronavirus in russia: The latest news, 2021{\natexlab{g}}.
\newblock URL
  \url{https://www.themoscowtimes.com/2021/02/01/coronavirus-in-russia-the-latest-news-feb-1-a69117}.

\bibitem[Times(2021{\natexlab{h}})]{F7}
The New~York Times.
\newblock France hired mckinsey to help in the pandemic. then came the
  questions., 2021{\natexlab{h}}.
\newblock URL
  \url{https://www.nytimes.com/2021/02/22/business/france-mckinsey-consultants-covid-vaccine.html}.

\bibitem[Times(2021{\natexlab{i}})]{Singapore7}
The~Straits Times.
\newblock Covid-19 vaccine and you: What to expect, 2021{\natexlab{i}}.
\newblock URL
  \url{https://www.straitstimes.com/singapore/covid-19-vaccine-and-you-what-to-expect}.

\bibitem[Times(2021{\natexlab{j}})]{Singapore8}
The~Straits Times.
\newblock All who take covid-19 jabs will get physical vaccination card, cannot
  choose which vaccine to take, 2021{\natexlab{j}}.
\newblock URL
  \url{https://www.straitstimes.com/singapore/politics/all-vaccinated-against-covid-19-will-get-physical-vaccination-card-cannot-choose}.

\bibitem[Today(2020)]{Japan17}
Business Today.
\newblock Covid-19 vaccine: Japan to spend \$6.3 billion from emergency budget
  reserve, 2020.
\newblock URL
  \url{https://www.businesstoday.in/sectors/pharma/covid-19-vaccine-japan-to-spend-63-billion-from-emergency-budget-reserve/story/415473.html}.

\bibitem[Today(2021{\natexlab{a}})]{Russ10}
Business Today.
\newblock Russia offers free transfer of covid-19 vaccine tech to india, other
  manufacturing partners, 2021{\natexlab{a}}.
\newblock URL
  \url{https://www.businesstoday.in/current/economy-politics/russia-offers-free-transfer-of-covid-19-vaccine-tech-to-india-other-manufacturing-partners/story/436620.html}.

\bibitem[Today(2021{\natexlab{b}})]{Eg2}
Egypt Today.
\newblock Egyptians aged above 40 yrs start registering for covid-19
  vaccination sunday, 2021{\natexlab{b}}.
\newblock URL
  \url{https://www.egypttoday.com/Article/1/99087/Egyptians-aged-above-40-yrs-start-registering-for-COVID-19}.

\bibitem[Today(2021{\natexlab{c}})]{Eg5}
Egypt Today.
\newblock Egypt vaccinates 149k individuals, to acquire total of 4.5m doses by
  june, 2021{\natexlab{c}}.
\newblock URL
  \url{https://www.egypttoday.com/Article/1/100476/Egypt-vaccinates-149K-individuals-to-acquire-total-of-4-5M}.

\bibitem[trackvaccines.org(2021)]{F3}
trackvaccines.org.
\newblock 4 vaccines approved for use in france, 2021.
\newblock URL \url{https://covid19.trackvaccines.org/country/france/}.

\bibitem[TRTWorld(2020)]{Turkey2}
TRTWorld.
\newblock How turkey plans to administer covid-19 vaccines, 2020.
\newblock URL
  \url{https://www.trtworld.com/magazine/how-turkey-plans-to-administer-covid-19-vaccines-42242}.

\bibitem[trust.org(2021)]{F10}
trust.org.
\newblock Covid-19: Are digital health passports a good idea?, 2021.
\newblock URL \url{https://news.trust.org/item/20201221125814-sv8ea/}.

\bibitem[UNHHCR(2021)]{Nepal4}
UNHHCR.
\newblock Nepal becomes first country in asia pacific to vaccinate refugees
  against covid-19, 2021.
\newblock URL
  \url{https://www.unhcr.org/news/stories/2021/3/6062c7184/nepal-becomes-first-country-asia-pacific-vaccinate-refugees-against-covid.html}.

\bibitem[Vaktsineeri(2021)]{Estonia2}
Vaktsineeri.
\newblock Vaccination in estonia, 2021.
\newblock URL \url{https://vaktsineeri.ee/en/covid-19/vaccination-in-estonia/}.

\bibitem[Valiati \& Villela(2021)Valiati and Villela]{valiati2021mitigation}
Naiara~CM Valiati and Daniel~AM Villela.
\newblock Mitigation policies and vaccination in the covid-19 pandemic: a
  modelling study.
\newblock \emph{medRxiv}, 2021.

\bibitem[Valitsus(2021)]{Estonia8}
Valitsus.
\newblock Estonia and the who signed a memorandum of understanding, 2021.
\newblock URL
  \url{https://www.valitsus.ee/en/news/estonia-and-who-signed-memorandum-understanding}.

\bibitem[vir.com.vn(2021)]{VT1}
vir.com.vn.
\newblock 11 priority groups eligible to first covid-19 vaccination in vietnam,
  2021.
\newblock URL
  \url{https://www.vir.com.vn/11-priority-groups-eligible-to-first-covid-19-vaccination-in-vietnam-82809.html}.

\bibitem[Wang et~al.(2020)Wang, Wu, Yang, Dong, Chen, Bai, Chen, Chen, Viboud,
  Ajelli, et~al.]{wang2020global}
Wei Wang, Qianhui Wu, Juan Yang, Kaige Dong, Xinghui Chen, Xufang Bai, Xinhua
  Chen, Zhiyuan Chen, C{\'e}cile Viboud, Marco Ajelli, et~al.
\newblock Global, regional, and national estimates of target population sizes
  for covid-19 vaccination: descriptive study.
\newblock \emph{bmj}, 371, 2020.

\bibitem[Website(2021)]{Singapore9}
Singapore Government~Agency Website.
\newblock Covid-19 vaccination, 2021.
\newblock URL \url{https://www.gov.sg/features/covid-19-vaccination}.

\bibitem[Wired.it(2021)]{It4}
Wired.it.
\newblock Come e quando prenotare il vaccino anti-covid-19, regione per
  regione, 2021.
\newblock URL
  \url{https://www.wired.it/scienza/medicina/2021/02/04/vaccino-covid-19-prenotazione/?refresh_ce}.

\bibitem[Xia et~al.(2021)Xia, Zhang, Wang, Wang, Yang, Gao, Tan, Wu, Xu, Lou,
  et~al.]{xia2021safety}
Shengli Xia, Yuntao Zhang, Yanxia Wang, Hui Wang, Yunkai Yang, George~Fu Gao,
  Wenjie Tan, Guizhen Wu, Miao Xu, Zhiyong Lou, et~al.
\newblock Safety and immunogenicity of an inactivated sars-cov-2 vaccine,
  bbibp-corv: a randomised, double-blind, placebo-controlled, phase 1/2 trial.
\newblock \emph{The Lancet Infectious Diseases}, 21\penalty0 (1):\penalty0
  39--51, 2021.

\bibitem[Xinhuanet(2021)]{Kaza2}
Xinhuanet.
\newblock Kazakhstan begins mass vaccination against covid-19, 2021.
\newblock URL
  \url{http://www.xinhuanet.com/english/2021-02/02/c_139713832_2.htm}.

\bibitem[Xpress(2021)]{Gm15}
Medical Xpress.
\newblock Germany reports cases of indian covid variant, 2021.
\newblock URL
  \url{https://medicalxpress.com/news/2021-04-germany-cases-indian-covid-variant.html}.

\bibitem[Zhang et~al.(2021)Zhang, Zeng, Pan, Li, Hu, Chu, Han, Chen, Tang, Yin,
  et~al.]{zhang2021safety}
Yanjun Zhang, Gang Zeng, Hongxing Pan, Changgui Li, Yaling Hu, Kai Chu, Weixiao
  Han, Zhen Chen, Rong Tang, Weidong Yin, et~al.
\newblock Safety, tolerability, and immunogenicity of an inactivated sars-cov-2
  vaccine in healthy adults aged 18--59 years: a randomised, double-blind,
  placebo-controlled, phase 1/2 clinical trial.
\newblock \emph{The Lancet Infectious Diseases}, 21\penalty0 (2):\penalty0
  181--192, 2021.

\end{thebibliography}
\bibliographystyle{iclr2021_conference}

\end{document}